\documentclass{article}
\pdfoutput=1 
\usepackage{jcappub}
\usepackage{amssymb}
\usepackage{amsmath}
\usepackage{hyperref}
\usepackage{graphicx}
\usepackage{caption}
\usepackage{subcaption}

\newcommand{\nhat}{\hat{\mathbf{n}}}
\newcommand{\phat}{\hat{\mathbf{p}}}
\newcommand{\qhat}{\hat{\mathbf{q}}}
\newcommand{\RR}{\mathcal{R}}
\newcommand{\lmaxrec}{\ell_{\max}^{\rm rec}}
\newcommand{\lmaxGWB}{\ell_{\max}^{\rm GWB}}

\definecolor{darkred}{RGB}{175,0,0}

\definecolor{darkteal}{RGB}{0,121,150}

\author[a,b]{Federico Semenzato,}
\emailAdd{federico.semenzato.1@phd.unipd.it}

\author[a,b,c]{Nicola Bellomo,}
\emailAdd{nicola.bellomo@unipd.it}

\author[a,b,c]{Alvise Raccanelli,}
\emailAdd{alvise.raccanelli.1@unipd.it}

\author[d]{Chiara M. F. Mingarelli}
\emailAdd{chiara.mingarelli@yale.edu}

\affiliation[a]{Dipartimento di Fisica e Astronomia ``G. Galilei'', Universit\`a degli Studi di Padova, via Marzolo 8, I-35131 Padova, Italy}
\affiliation[b]{INFN, Sezione di Padova, Via Marzolo 8, I-35131, Padova, Italy}
\affiliation[c]{INAF - Osservatorio Astronomico di Padova, Vicolo dell'Osservatorio 5, I-35122 Padova, Italy}
\affiliation[d]{Department of Physics, Yale University, New Haven, CT, 06520, USA}

\title{Bias from small-scale leakage in Pulsar Timing Array maps}

\abstract{
Pulsar Timing Array experiments are rapidly approaching the era of gravitational wave background anisotropy detection.
The timing residuals of each pulsar are an integrated measure of the gravitational-wave power across all angular scales. 
However, due to the limited number of monitored pulsars, current analyses are only able to reconstruct the angular structure of the background at large scales.
We show analytically that this mismatch between the integrated all-sky signal and the truncated reconstruction introduces a previously unaccounted source of systematic bias in the anisotropic background angular power spectrum.
The source of this systematic error, that we call ``small-scale leakage'', is the intrinsic presence of unaccounted gravitational wave power at scales smaller than the reconstructed scales.
This unmodeled power leaks into large-scale modes, artificially increasing the recovered value of the inferred angular power spectrum by at least one order of magnitude in a wide range of scales.
Importantly, this effect is fundamentally independent of the geometry of the pulsar configuration, the anisotropy reconstruction method, the use of different regularization schemes, and the presence of pulsar noise.
As the quality of pulsar timing array experiments improves, a robust understanding of small-scale leakage will become paramount for reliable detection and characterization of the gravitational wave background.
Thus, the theoretical formalism developed here will be essential to estimate the magnitude of this systematic uncertainty in anisotropy searches. 
}

\begin{document}

\maketitle

\section{Introduction}

Anisotropies of the nanohertz gravitational wave background (GWB) contain invaluable information regarding its origin, whether astrophysical~\cite{sesana_stochastic_2008,Agazie_2023_smbhb,satopolito2023nanogravsbigblackholes,sah2024_imprints}, cosmological~\cite{Starobinsky:1979ty,SIGW,Afzal_2023} or a mixture of both.
At this time, numerous Pulsar Timing Array (PTA) collaborations already reported evidence for a GWB~\cite{agazie_nanograv_2023, eptacollaboration_second_2023, reardon_search_2023, xu_searching_2023, Miles_2024}.
However, despite a robust understanding of the capabilities of detecting its anisotropies~\cite{Mingarelli_2013, Taylor_2013, Mingarelli_2017, Gardiner2024, Moreschi25}, we have just started to gather hints about their presence~\cite{agazie2023_anisotropies, Grunthal_2024}. 
Their detection opens up the possibility of locating individual supermassive black hole binaries (SMBHBs) in well-localized host galaxies~\cite{Mingarelli_2017, Casey_Clyde_2022, agazie2024_ACC, tian2025targetedsearchindividualsmbhb, agarwal2025nanograv15yrdata, CaseyClyde2025, Schult_25} and, for future experiments, to cross-correlate the anisotropic GWB with galaxy surveys to firmly establish its origin~\cite{Bellomo_2022, semenzato2024crosscorrelatinguniversegravitationalwave, sah2024, cusin25}. 

However, measuring and interpreting the statistics of the GWB anisotropies still represents a formidable challenge~\cite{pol_taylor, konstandin2024impactcosmicvarianceptas, konstandin2025prospectslimitationsptasanisotropy, Gersbach25, Bernardo25, domcke2025cosmicvarianceanisotropysearches}. 
Any PTA is comprised of a finite and sparsely distributed set of millisecond pulsars, and each pulsar timing residual is generated by the integrated GW power distributed across the entire sky.
Thus, because of the limited number of pulsar pairs, we can reconstruct the GWB angular distribution only at the largest angular scales~\cite{alihaimoud2020, taylor2020,domcke2025cosmicvarianceanisotropysearches}. 
On the other hand, the actual GW sky, potentially shaped by a large number of SMBHBs, contains a structure that extends to rather small angular scales, which appears to be discarded from traditional analyses~\cite{Mingarelli_2017, agazie2024_ACC}.

Although it is not possible to draw any perfect analogy with the case discussed in this work, we note that cosmology has already dealt with potential issues regarding the interpretation of information coming from small-scales, in particular in the realm of simulation and data analysis.
In the case of Cosmic Microwave Background, two instances immediately come to mind: the need to remove small-scale power at scales larger than the desired resolution of the analysis to avoid aliasing effects~\cite{Hu_2003, Alonso_2019}, and the need to remove small-scale power when downgrading an existing map to a lower resolution~\cite{sullivan2024methodscmbmapanalysis}.
Failure to perform such removal results in both cases in overestimating the magnitude of anisotropies or, equivalently, of their statistical $n$-point functions.
Within the existing Cosmic Microwave Background literature, this contamination of the large-scale angular power spectrum by unresolved small-scale power is a well-studied problem.
For instance, refs.~\cite{Tegmark:1996qt,Bond:1998zw} discuss how in an optimal quadratic estimator framework it is possible to deal with unresolved modes by explicitly including them in the data covariance.
The key insight in this instance, which has been fully formalized by ref.~\cite{Seljak:1997wx}, is that when the data contain signal from modes that are not being estimated, the covariance acquires an additional ``aliasing'' contribution originating from those unmodeled modes.
If this aliasing contribution is omitted from the analysis, the resulting estimates turn out to be biased.

Additionally, it is well understood how aliasing effects appear when analyzing dark matter or galaxy catalogs to extract their power spectrum due to the finite resolution of the Fourier mode grid~\cite{nbodybook, Sefusatti_2016, Hand_2018}.
As in the previous case, the consequence is a conspicuous overestimation of the two-point function.
The saving grace in all these instances is the possibility to directly access the map of spatial anisotropies, and to perform some form of ``spatial smoothing'' that removes small-scale power before performing any statistical analysis.

In a PTA, GWB anisotropies contribute to each correlated timing residual as an integrated signal over the sky. 
Consequently, power on small angular scales (although unreconstructable) is still embedded in every residual measurement. 
When the GWB sky is modeled only up to a finite multipole $\ell_{\max}$, this unresolved small-scale power is projected onto the available large-scale modes. 
As a result, forcing all measured power into low-$\ell$ modes inevitably leads to an overestimation of the anisotropy amplitude and a distorted angular power spectrum.

We call this phenomenon \textit{small-scale leakage}: it is a systematic bias that has not yet been properly investigated in the PTA context. 
The relevance of this effect is expected to become a fatal source of error in any anisotropy measurement, as PTA collaborations rapidly improve the precision of timing residual measurements, the number of monitored pulsars, and the total observation time. 
Therefore, without a rigorous understanding of this systematic bias and the development of effective mitigation strategies, this could emerge as an important challenge for future claims about the detection of anisotropies and their physical interpretation.

This work presents the first complete analytical framework for understanding and quantifying small-scale leakage in PTA anisotropy measurements. 
First, we demonstrate the presence and magnitude of small-scale power in timing residuals measurements.
Then, we derive analytical expressions describing how unmodeled power propagates through the reconstruction process, revealing the large-scale modes contamination sourced by small-scale power.
Furthermore, we explicitly show which interplay exists between small-scale leakage and regularization strategies implemented in data analyses. 
These results provide the theoretical foundation for the development of statistical estimators that account for such a systematic effect, thus allowing for the reconstruction of uncontaminated large-scale GWB sky maps.

This paper is organized as follows. 
Section~\ref{sec:pta_response} provides a quick overview of the basic concepts used in this work.
Section~\ref{sec:smallscale_leakage} demonstrates the existence of small-scale leakage bias and its interplay with regularization schemes in the idealized case of noiseless measurements. 
Section~\ref{sec:pulsar_noise} shows how the small-scale leakage effect persists even when noisy measurements are considered. 
Finally, we conclude in section~\ref{sec:discussion_conclusions} with a discussion of the implications of our findings for upcoming anisotropy searches.
The appendices~\ref{app:persistence_smallscale_residuals}, \ref{app:smallscale_leakage_complementary}, and~\ref{app:regularization} contain additional material that supports the discussion in the main text.
In this work, we adopt natural units $c=G=1$.
A summary of our notation is provided in table~\ref{tab:notation}.

\begin{table}[h]
    \centering
    \renewcommand{\arraystretch}{1.15}
    \footnotesize
    \begin{tabular}{|c|c|c|}
    \hline
    \textbf{Symbol} & \textbf{Description} & \textbf{Defining equation} \\
    \hline\hline
    $R_p$ & Frequency-space timing residuals for pulsar \textit{p} & \eqref{eq:Rpf} \\
    \hline
    $\mathcal{R}_{pq}$ & Correlated residual for pulsar pair $(p,q)$ & \eqref{eq:timing_residuals_twopointfunction} \\
    \hline
    $\boldsymbol{\RR}$ & Vector of GW-induced correlated residuals for all pulsar pairs & \eqref{eq:map_making} \\
    \hline
    $\boldsymbol{\RR}^\mathrm{meas}$ & Vector of measured correlated residuals for all pulsar pairs & \eqref{eq:map_making} \\
    \hline
    $\boldsymbol{\RR}_{\mathrm{LS}},\ \boldsymbol{\RR}_{\mathrm{SS}}$ & Large/Small-Scale correlated residual & \eqref{eq:leakage_existence} \\
    \hline
    $\Gamma,\ \Gamma_{\mathrm{LS}},\ \Gamma_{\mathrm{SS}}$ & Geometry matrix (full; LS/SS blocks) & \eqref{eq:geometry_kernel}, \eqref{eq:leakage_existence} \\
    \hline
    $\mathbf{a}_{\mathrm{LS}}$ & LS harmonic coefficients estimator & \eqref{eq:shift_least_square}, \eqref{eq:shift_maximum_likelihood} \\
    \hline
    $\mathbb{P}_\ell$ & Projection matrix selecting the $(2\ell+1)$ LS block & \eqref{eq:cl_estimator}, \eqref{eq:Pell_def} \\
    \hline
    $M_{\ell\ell'}, M'_{\ell\ell'}, M''_{\ell\ell'}$ & Mode-mixing leakage kernel & \eqref{eq:mode_mixing_pure}, \eqref{eq:cl_leakage_likelihood}, \eqref{eq:Msec_cell}
    \\
    \hline
    $\mathrm{Cov},\ \mathrm{Cov}_{\mathrm{reg}}$ & Residual covariance and SVD-regularized version & \eqref{eq:cov_puresignal}, \eqref{eq:full_cov} \\
    \hline
    $\lambda,\ \mathcal{P}$ & Ridge parameter and penalty matrix & \eqref{eq:maximumlikelihoodshift_regularization} \\
    \hline
    $K_{\ell\ell'}$ & Mode-mixing  regularization-bias kernel & \eqref{eq:reg_bias_cell} \\
    \hline
    $\mathcal{H},\ \mathbf{v}$ & Likelihood Hessian and score & \eqref{eq:normal equations} \\
    \hline
    $e_{\mathrm{thres}}$ & SVD eigenvalue cutoff & \eqref{eq:SVD_def} \\
    \hline
    \end{tabular}
    \caption{Summary of the notation used in this paper in order of appearance.}
    \label{tab:notation}
\end{table}

%%%%%%%%%%%%%%%%%%%%%%%%%%%%%%%%%%%%%%%%%%%%%%%%%%%%%%%%%%%%%%%%%%%%%%%%%%%%%%%%%%%%%%%%%%%%%%%%%%%%%%%%%%%%%%%%%%%%%%%%%%%%%%%%%%%%

\section{Pulsar timing array response to a gravitational wave background}
\label{sec:pta_response}

The metric fluctuation~$h_{ij}$ describing the GWB is usually decomposed in individual plane waves as
\begin{equation}
    h_{ij}(t,\mathbf{x}) = \sum_A \int_{-\infty}^{\infty} df \int d\nhat \ h_A(f,\nhat) e^{A}_{ij}(\nhat) e^{i 2\pi f (t-\nhat\cdot\mathbf{x})},
\end{equation}
where~$(t,\mathbf{x})$ are the cosmic time and comoving coordinates, respectively, $(f,\nhat)$ are the GW frequency and propagation direction, respectively, $A=\{+,\times\}$ are the two polarization degrees of freedom, $e^{A}_{ij}(\nhat)$ is the polarization tensor, and~$h_A(f,\nhat)$ is the mode amplitude.
In the case of a stationary, Gaussian, and unpolarized background, we have
\begin{equation}
    \left\langle h_A \left(f, \nhat\right) h^*_{A'} \left( f', \nhat' \right) \right\rangle_s = \delta^K_{AA'} \delta^D \left( f-f' \right) \delta^D \left(\nhat-\nhat'\right) \frac{H(f,\nhat)}{8\pi},
\end{equation}
where~$\left\langle\ \cdot\ \right\rangle_s$ indicates the ensemble average over the properties of the population of sources (masses, spins, eccentricities, etc.), $^*$ indicates the complex conjugate, $\delta^K$ and~$\delta^D$ are Kronecker and Dirac deltas, respectively, and~$H(f,\nhat)$ is the (one-sided) GWB power spectrum.
In this work, we consider a GWB produced by a large number of SMBHBs stochastically distributed in host galaxies.
Therefore, since their spatial distribution is intrinsically anisotropic, the GWB power spectrum retains a dependence on the direction.
Although in this work we are interested in the ``weak field'' regime, where the anisotropies in the emitted power in GWs are expected to be small since they are not generated by individual ``loud sources'', the effect discussed in this paper exists for every form of the GWB power spectrum.
Under these assumptions, we can conveniently factorize the power spectrum as
\begin{equation}
    H(f,\nhat) = H(f) \left[ 1 + P(f,\nhat) \right],
\end{equation}
where~$H(f)$ is the isotropic value, $P(f,\nhat)$ characterizes the anisotropic angular distribution of GWB, and it has zero mean across the entire sky, i.e., $\displaystyle \int d\nhat\ P(f,\nhat) = 0$. 
In this notation, the physically meaningful values of the anisotropic power are~$P(f,\nhat) \geq -1$.
Since in this work we consider a GWB sourced by a large population of SMBHBs, it is appropriate to consider the ``GW sky'' as time-invariant since the collective emission pattern does not vary on the observationally relevant timescales.

The presence of a GWB causes a delay in the arrival time of pulsar electromagnetic signals.
The induced, frequency-dependent timing residual in the arrival time from a pulsar~$p$ at distance~$D_p$ in the direction~$\phat$ reads as
\begin{equation}
    R_p(f) = \frac{1}{2\pi i f} \int d\nhat \sum_A h_A(f,\nhat) F^A_p(\nhat) \left[ 1 - e^{-2\pi i f D_p (1+\nhat\cdot\phat)} \right],
\label{eq:Rpf}
\end{equation}
where the pulsar antenna pattern function is given by
\begin{equation}
    F^A_p(\nhat) = \frac{\hat{p}^i \hat{p}^j}{2\,(1+\nhat\cdot\phat)} e^A_{ij}(\nhat).
\label{eq:pattern}
\end{equation}
Therefore, the expectation value of the timing residual coming for a pair of different pulsars~$(p,q)$ reads as
\begin{equation}
     \left\langle R_p(f) R_q^*(f') \right\rangle_s = \frac{\delta^D(f-f')}{2} \left[ \RR^\mathrm{iso}_{pq} + \RR^\mathrm{anis}_{pq} \right],
\end{equation}
where we define as ``isotropic'' and ``anisotropic'' correlated residuals those contributions sourced by the isotropic and anisotropic components of the GWB, respectively. 
They are formally defined as
\begin{equation}
    \begin{aligned}
        \RR^\mathrm{iso}_{pq} &= \frac{H(f)}{(2\pi f)^2} \int \frac{d\nhat}{4\pi} \gamma_{pq}(\nhat) = \frac{H(f)}{(2\pi f)^2} \mathrm{HD}(\phat \cdot \qhat),  \\
        \RR^\mathrm{anis}_{pq} &= \frac{H(f)}{(2\pi f)^2} \int \frac{d\nhat}{4\pi} P(f,\nhat) \gamma_{pq}(\nhat), \\
    \end{aligned}
\label{eq:timing_residuals_twopointfunction}
\end{equation}
where the overlap reduction functions read as
\begin{equation}
    \gamma_{pq}(\nhat) =  \sum_A F^A_p(\nhat) F^A_q(\nhat) \left[ 1 - e^{-2\pi i f D_p (1+\nhat \cdot \phat)} \right] \left[ 1 - e^{2\pi i f D_q (1+\nhat \cdot \qhat)} \right] \approx \sum_A F^A_p(\nhat) F^A_q(\nhat),
\label{eq:gamma-kernel}
\end{equation}
since the Earth–pulsar phase factors in square brackets give a rapidly oscillatory contribution that can be neglected in the long-arm limit~\cite{MM2018}, and
\begin{equation}
    \mathrm{HD}(\phat \cdot \qhat) = \int \frac{d\nhat}{4\pi} \gamma_{pq}(\nhat) = \frac{1}{3} - \frac{1-\phat \cdot \qhat}{12} + \frac{1-\phat \cdot \qhat}{2}\log \frac{1-\phat \cdot \qhat}{2}
\end{equation}
is the Hellings-Downs curve~\cite{HD83}.\footnote{
Multiple definitions of Hellings-Downs are currently used in the literature. 
Our definition is related to the one employed by the NANOGrav Collaboration by a rescaling factor as~$\mathrm{HD}_\mathrm{NANOGrav}(\phat \cdot \qhat) = (3/2) \mathrm{HD}(\phat \cdot \qhat)$.}

In general, the measured pulsar timing residual~$R^\mathrm{meas}_p$ is given by two contributions, i.e.
\begin{equation}
    R^\mathrm{meas}_p = R_p + n_p,
\label{eq:measured_timing_residual}
\end{equation}
where~$R_p$ is the residual defined in equation~\eqref{eq:Rpf}, and the noise~$n_p$ is assumed to have zero mean~$\left\langle n_p \right\rangle = 0$ and variance~\cite{alihaimoud2020, Ali_Ha_moud_2021}
\begin{equation}
    \left\langle n_p n_q \right\rangle_n = \frac{\delta^{D}(f-f')}{2} \delta^K_{pq} \sigma^2_p(f),
\end{equation}
where~$\left\langle\ \cdot\ \right\rangle_n$ indicates ensemble averages over noise realizations, $\sigma^2_p(f)$ is the (one-sided) noise power spectrum, and we implicitly assume that the pulsar noise is uncorrelated between different pulsars and is also uncorrelated with the GWB signal, i.e., $\left\langle R_p n_q \right\rangle = 0$.

%%%%%%%%%%%%%%%%%%%%%%%%%%%%%%%%%%%%%%%%%%%%%%%%%%%%%%%%%%%%%%%%%%%%%%%%%%%%%%%%%%%%%%%%%%%%%%%%%%%%%%%%%%%%%%%%%%%%%%%%%%%%%%%%%%%%

\section{Emergence of small-scale leakage}
\label{sec:smallscale_leakage}

For the sake of clarity, we demonstrate the existence of a purely physical effect called ``small-scale leakage'' in the case of noiseless timing residual measurements.
Pulsar noise is introduced in section~\ref{sec:pulsar_noise}, alongside with a discussion of its impact on small-scale leakage.

%%%%%%%%%%%%%%%%%%%%%%%%%%%%%%%%%%%%%%%%%%%%%%%%%%%%%%%%%%%%%%%%%%%%%%%%%%%%%%%%%%%%%%%%%%%%%%%%%%%%%%%%%%%%%%%%%%%%%%%%%%%%%%%%%%%%

\subsection{Map-making process}

Different clustering properties of GW-emitting sources are expected to have a different impact on the magnitude of anisotropic correlated timing residuals.
Since these properties are typically scale dependent, it is convenient to decompose the GWB power anisotropic distribution appearing in equation~\eqref{eq:timing_residuals_twopointfunction} in a basis that makes explicit how different scales contribute to the correlated residual.
The considerations raised in this article are independent from the particular choice of basis; however, in order ot present a fully worked-out scenario, here we adopt a real spherical harmonic decomposition.
In this basis, the GWB power angular distribution reads as
\begin{equation}
    P(f,\nhat) = \sum_{\ell m} a^f_{\ell m} \mathcal{Y}_{\ell m}(\nhat)\,,
\label{eq:real_spherical_harmonics_decomposition}
\end{equation}
where~$\mathcal{Y}_{\ell m}$ are the real spherical harmonics, and~$a^f_{\ell m}$ are the frequency-dependent harmonic coefficients.\footnote{
Real spherical harmonics are defined in terms of spherical harmonics~$Y_{\ell m}$ as~$\mathcal{Y}_{\ell m} = \sqrt{2} \mathrm{Re} (Y_{\ell m})$ ($m>0$), $\mathcal{Y}_{\ell 0} = Y_{\ell 0}$ ($m=0$), and~$\mathcal{Y}_{\ell m} = \sqrt{2} \mathrm{Im} (Y_{\ell |m|})$ ($m<0$).}
These coefficients represent a true random variable associated with the stochasticity of the position of the sources.
In other words, even considering a fixed set of GW sources, the scenario presents a residual degree of stochasticity given by how GW sources populate different host galaxies.
The mean and variance of the harmonic coefficients read as
\begin{equation}
    \left\langle a^f_{\ell m} \right\rangle = 0, \qquad \left\langle a^f_{\ell m} a^{f'}_{\ell' m'} \right\rangle = \delta_{\ell\ell'} \delta_{m m'} C^{ff'}_\ell,
\label{eq:alm_stats}
\end{equation}
where the angular power spectrum~$C^{ff'}_\ell \neq 0$ even when~$f \neq f'$ when the background is sourced by eccentric SMBHBs~\cite{moreschi2025dissectingnanohzgravitationalwave, sah2025accuratemodelingnanohertzgravitational}, and, more generally, because the spatial distribution of the hosts of GW sources emitting in different frequency bands is correlated.
The angle brackets~$\left\langle\ \cdot\ \right\rangle$ appearing in equation~\eqref{eq:alm_stats} indicate averaging over the~$a_{\ell m}$ ensemble, i.e., over all the possible spatial distributions of the same SMBHB population across different galaxies.
Finally, from now on, we consider only multipoles~$\ell \geq 2$ because~$P(f,\nhat)$ has no isotropic component by construction and, in the weak field regime, the value of the harmonic coefficients with~$\ell=1$ is completely dominated by the kinetic dipole.

Recovering the angular power spectrum that characterizes the anisotropic distribution of GW sources is tightly connected to our ability to reconstruct the~$P(f,\nhat)$ map.
Suppose to monitor a set of~$N_\mathrm{psr}$ pulsars, and to have obtained a set of anisotropic correlated timing residual for each one of the~$N_\mathrm{pairs}=N_\mathrm{psr}(N_\mathrm{psr}-1)/2$ pulsar pairs of the sample.
Consider also a single frequency bin for simplicity.
Data can then be organized in a vectorial structure as~\cite{taylor2020, alihaimoud2020, Ali_Ha_moud_2021}
\begin{equation}
    \begin{aligned}
        & \boldsymbol{\RR}^\mathrm{meas} = \boldsymbol{\RR} + \mathbf{n} = 
        \left( \begin{matrix}
            \RR^\mathrm{anis}_{p_1 p_2} \\
            \RR^\mathrm{anis}_{p_1 p_3} \\
            \vdots \\
            \RR^\mathrm{anis}_{p_{N-1} p_N}
        \end{matrix} \right) + 
        \left( \begin{matrix}
            n_{p_1 p_2} \\
            n_{p_1 p_3} \\
            \vdots \\
            n_{p_{N-1} p_N}
        \end{matrix} \right) \\
        &=
        \left( \begin{matrix}
            \Gamma^{p_1 p_2}_{2\; -2} & \Gamma^{p_1 p_2}_{2\; -1} & \Gamma^{p_1 p_2}_{2\; 0} & \dots & \Gamma^{p_1 p_2}_{\ell m} & \dots &  \\
            \vdots & & \vdots & & \vdots & & \\
            \vdots & & \vdots & & \vdots & & \\
            \Gamma^{p_{N-1} p_N}_{2\; -2} & \Gamma^{p_{N-1} p_N}_{2\; -1} & \Gamma^{p_{N-1} p_N}_{2\; 0} & \dots & \Gamma^{p_{N-1} p_N}_{\ell m} & \dots &  \\
        \end{matrix} \right) 
        \left( \begin{matrix}
            a_{2\; -2} \\
            a_{2\; -1} \\
            a_{2\; 0} \\
            \vdots \\
            a_{\ell m} \\
            \vdots 
        \end{matrix} \right) +
        \left( \begin{matrix}
            n_{p_1 p_2} \\
            n_{p_1 p_3} \\
            \vdots \\
            n_{p_{N-1} p_N}
        \end{matrix} \right)
        = \Gamma \mathbf{a} + \mathbf{n},
    \end{aligned}
\label{eq:map_making}
\end{equation}
where~$\boldsymbol{\RR}^\mathrm{meas}, \boldsymbol{\RR}, \mathbf{n}$ are~$(N_\mathrm{pairs} \times 1)$ column vectors containing the measured correlated residuals, and its decomposition into a signal a noise component, respectively, the elements of the~$\Gamma$ matrix, using equations~\eqref{eq:timing_residuals_twopointfunction} and~\eqref{eq:real_spherical_harmonics_decomposition}, read as
\begin{equation}
    \Gamma^{p_i p_j}_{\ell m} = \frac{H(f)}{(2\pi f)^2} \int \frac{d\nhat}{4\pi} \gamma_{p_i p_j}(\nhat) \mathcal{Y}_{\ell m}(\nhat),
    \label{eq:geometry_kernel}
\end{equation}
and~$\mathbf{a}$ is a column vector that contains the harmonic coefficients.
Since it is not known a priori how the anisotropic power is distributed on different scales, the dimension of the~$\mathbf{a}$ vector, and thus the number of columns of the~$\Gamma$ matrix, has the potential of being extremely large.
On the other hand, we probe a finite amount of pulsar pairs, therefore we have the ability of constraining at most~$N_\mathrm{pairs}$ real spherical harmonic coefficients or, equivalently, multipoles up to a maximum reconstructed scale~$\lmaxrec = \left\lfloor \sqrt{\frac{N_\mathrm{psr}(N_\mathrm{psr}-1)}{2} + 4} -1 \right\rfloor$~\cite{alihaimoud2020, domcke2025cosmicvarianceanisotropysearches}, contrariwise to what reported in ref.~\cite{Romano_2017}. 

Regarding the structure of the~$\mathbf{a}$ vector, suppose that we have a GWB with anisotropic power up to some maximum scale~$\lmaxGWB > \lmaxrec$, i.e., that~$a_{\ell m}=0$ for every multipole~$\ell>\lmaxGWB$.
In this scenario, we can divide the set of harmonic coefficients into a set of ``large-scale'' (LS) coefficients with multipoles~$2\leq \ell \leq \lmaxrec$, and a set of ``small-scale'' (SS) coefficients with multipoles~$\lmaxrec+1 \leq \ell \leq \lmaxGWB$, that we then arrange in two vectors as
\begin{equation}
    \mathbf{a}_\mathrm{LS} = 
    \left( \begin{matrix}
        a_{2\; -2} \\
        a_{2\; -1} \\
        \vdots \\
        a_{\lmaxrec\; \lmaxrec-1} \\
        a_{\lmaxrec\; \lmaxrec}
    \end{matrix} \right), 
    \qquad\qquad
    \mathbf{a}_\mathrm{SS} = 
    \left( \begin{matrix}
        a_{\lmaxrec+1\;-(\lmaxrec+1)} \\
        a_{\lmaxrec+1\;-(\lmaxrec+1)+1} \\
        \vdots \\
        a_{\lmaxGWB\; \lmaxGWB-1} \\
        a_{\lmaxGWB\; \lmaxGWB}
    \end{matrix} \right),
\end{equation}
where the dimension of each vector is
\begin{equation}
    N_\mathrm{LS} = \mathrm{dim}(\mathbf{a}_\mathrm{LS}) = \left(\lmaxrec + 1\right)^2 - 4, \qquad N_\mathrm{SS} = \mathrm{dim}(\mathbf{a}_\mathrm{SS}) = \left(\lmaxGWB + 1\right)^2 - \left(\lmaxrec + 1\right)^2,
\end{equation}
respectively.
Under these assumptions, equation~\eqref{eq:map_making} nicely factorizes as 
\begin{equation}
\begingroup
  \setlength{\fboxrule}{1pt}% border thickness (default ~0.4pt)
  \setlength{\fboxsep}{6pt}% padding inside the box
  \scalebox{1.}{\(\boxed{\boldsymbol{\RR}^\mathrm{meas} = \boldsymbol{\RR}_\mathrm{LS} + \boldsymbol{\RR}_\mathrm{SS} + \mathbf{n}
  = \Gamma_\mathrm{LS}\mathbf{a}_\mathrm{LS} + \Gamma_\mathrm{SS}\mathbf{a}_\mathrm{SS} + \mathbf{n} }\)}
\endgroup
\label{eq:leakage_existence}
\end{equation}
where~$\Gamma_\mathrm{LS}$ contains the first~$\lmaxrec-1$ columns of~$\Gamma$, while~$\Gamma_\mathrm{SS}$ contains the second~$\lmaxGWB-\lmaxrec$ ones.
For later convenience, we define the correlated residuals coming from large- and small-scale modes only as~$\boldsymbol{\RR}_\mathrm{LS}$ and~$\boldsymbol{\RR}_\mathrm{SS}$, respectively.

The core idea of this work is contained in equation~\eqref{eq:leakage_existence}.
Correlated timing residuals generated by the anisotropic distribution of sources are integrated across all scales, even if limited samples of pulsars allow for the reconstruction of only a limited subset of harmonic coefficients.
Therefore, the extra small-scale power is unavoidably incorporated into the large-scale harmonic coefficients when we interpret the total correlated residuals as a large-scale only contribution by reconstructing the map up to~$\lmaxrec<\lmaxGWB$.
In other words, small-scale power leaks into large-scale modes when we attempt to perform a partial reconstruction of the anisotropic GWB power map, artificially increasing large-scale mode variance, i.e., their~$C_\ell$s.

\begin{figure}[ht]
    \centerline{
    \includegraphics[width=\columnwidth]{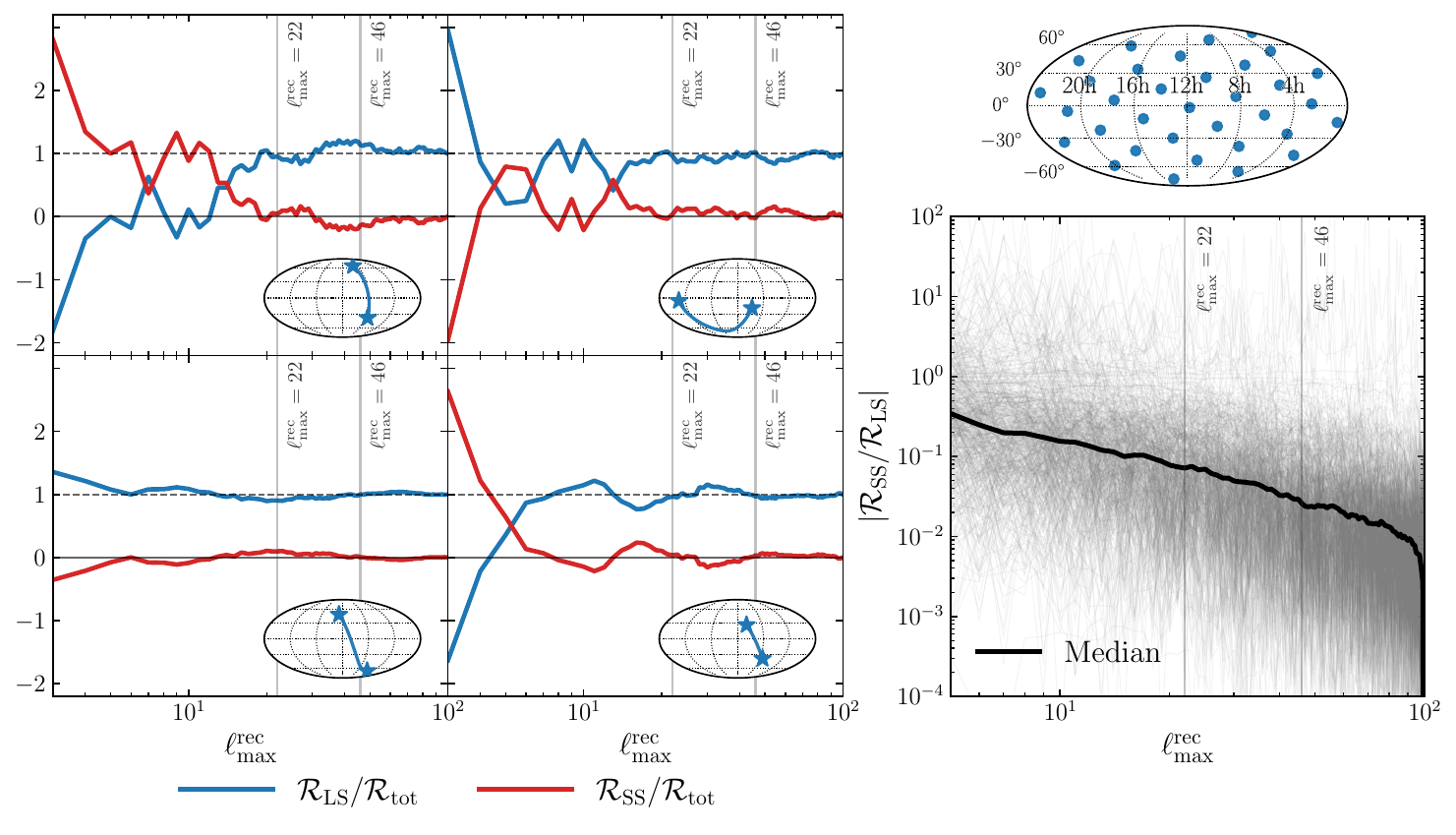}}
    \caption{\textit{Left panels}: fractional contribution of large-scale (\textit{blue lines}) and small-scale (\textit{red lines}) correlated residuals for four randomly displaced pulsars as a function of the reconstruction multipole~$\lmaxrec$. 
    For~$\lmaxrec=0$ (monopole only), the entire signal originates from small scales. 
    As~$\lmaxrec$ increases, progressively more power shifts into the large-scale component. 
    The sum of both components, i.e., the total correlated residual, remains constant when varying~$\lmaxrec$.
    \textit{Right panel}: absolute value of the small- to large-scale correlated residuals ratio for all pulsar pairs in the I34 configuration.
    The \textit{black line} represents the median of the absolute value of the ratio.}
\label{fig:RLS_RSS_I34example}
\end{figure}

Although the systematic bias effect presented in this work holds for every injected GWB, our proof-of-principle analysis considers exclusively the scenario where the GWB is sourced by a population of SMBHBs spatially distributed in galaxies, i.e., of a GWB with anisotropies resembling those of galaxy distributions.
Anisotropic maps of the GWB are generated using the~$10^3$ realizations and the pipeline created in ref.~\cite{semenzato2024crosscorrelatinguniversegravitationalwave}, and are compatible with the scenario in which the GWB is generated by a population of SMBHBs emitting at~$4~{\rm nHz}$, with no loud continuous-wave source present in the signal. 
This set of realizations is compatible with the weak-field regime assumption, and the inferred~$C_\ell \simeq 3\times 10^{-4}$ follows the expected Wishart distribution~\cite{semenzato2024crosscorrelatinguniversegravitationalwave}.
Despite the fact that GWB maps created with this approach presents power even at scales~$\ell \gtrsim 10^3$, for the sake of reducing computational cost, we consider maps where we truncate it at~$\lmaxGWB=250$.
In other words, all the results should to be interpreted as conservative, since there is considerably more power at small-scales than the one used to perform this analysis.
Regarding the pulsar distribution, we consider three sets of geometries with a different number of pulsars to showcase the robustness of the physical effect discussed in this article with respect to observational limitations.
These configurations are labeled as
\begin{itemize}
    \item \textbf{I34}, \textbf{I68}: two sets of~$N_\mathrm{psr}=34$ and~$N_\mathrm{psr}=68$ pulsars \textit{isotropically} distributed on the sky, respectively;
    \item \textbf{U34}, \textbf{U68}: two sets of~$N_\mathrm{psr}=34$ and~$N_\mathrm{psr}=68$ pulsars \textit{uniformly} distributed on the sky, respectively;
    \item \textbf{NG34}, \textbf{NG68}: a subset of~$N_\mathrm{psr}=34$ 15-year dataset NANOGrav pulsars, and the full sample.
\end{itemize}
Given these numbers of pulsars, the maximum reconstruction multipole is~$\lmaxrec=22$ ($\lmaxrec=46$) for the~$\{ \mathrm{I34, U34, NG34} \}$ ($\{\mathrm{I68, U68, NG68} \}$) configurations.

In the left panel of figure~\ref{fig:RLS_RSS_I34example} we show the relative importance of the small- and large-scale coefficients for four random pairs of pulsars as we change the maximum reconstruction multipole.
Although in the limit of large~$\lmaxrec$ the small-scale residuals carry a negligible amount of power, the same cannot be said for~$\lmaxrec \lesssim 10$, where the the large pile-up of small-scale power is clearly exacerbated.
These examples are not peculiar, as we show in the right panel of the same figure, where we report the ratio of small- to large-scale correlated residuals for all the pulsar pairs of the I34 configuration.
In particular, we observe that small- and large-scale residuals easily have the same order of magnitude for plausible values of the maximum reconstruction multipole.
Additionally, the broad dispersion of the small-scale-to-total correlated residuals shown in the right panel of figure~\ref{fig:RLS_RSS_I34example} already indicates the possibility that unresolved small-scale power can be imprinted in a structured fashion in the pair data, rather than behaving only as a single featureless contribution.
It also highlights the fact that different pulsar pairs have different angular separations and therefore different effective sensitivity to unresolved structure.
Moreover, these considerations are fundamentally independent of the pulsar geometry considered, as explicitly shown in appendix~\ref{app:persistence_smallscale_residuals} for the other pulsar configurations.

%%%%%%%%%%%%%%%%%%%%%%%%%%%%%%%%%%%%%%%%%%%%%%%%%%%%%%%%%%%%%%%%%%%%%%%%%%%%%%%%%%%%%%%%%%%%%%%%%%%%%%%%%%%%%%%%%%%%%%%%%%%%%%%%%%%%

\subsection{Small-scale leakage - Geometrical intuition}
\label{subsec:smallscaleleakage_geometricalintuition}

There are multiple fashions in which it is possible to establish the consequences of assigning small-scale power to large-scale modes.
Since this is the first time this concept is discussed in the PTA literature, we take a step-by-step approach to build up physical intuition.
In the next sections, we will refine this idea by introducing a likelihood approach, regularization schemes, and pulsar noise.

A first intuitive approach is to view the issue as a purely geometrical map-making process, where we attempt to directly reconstruct the GWB anisotropic map by implementing a ``minimum least squares'' estimator.
Suppose to look for a set of large-scale only correlated residuals that match the ``true'' ones~$\boldsymbol{\RR}$ via the idealized, noiseless ($\mathbf{n}=\mathbf{0}$) chi-squared statistics
\begin{equation}
    \chi^2_\mathrm{LMS} = \left| \boldsymbol{\RR} - \Gamma_\mathrm{LS} \mathbf{a}_\mathrm{LS} \right|^2 = \left| \Gamma_\mathrm{LS} \mathbf{a}^\mathrm{true}_\mathrm{LS} + \Gamma_\mathrm{SS} \mathbf{a}^\mathrm{true}_\mathrm{SS} - \Gamma_\mathrm{LS} \mathbf{a}_\mathrm{LS} \right|^2.
\label{eq:least_square}
\end{equation}
It is commonly known that the solution to this minimization problem is given by
\begin{equation}
\begingroup
  \setlength{\fboxrule}{1pt}% border thickness (default ~0.4pt)
  \setlength{\fboxsep}{6pt}% padding inside the box
  \scalebox{1.}{\(\boxed{\mathbf{a}_\mathrm{LS} = \mathbf{a}^\mathrm{true}_\mathrm{LS} + \Gamma^+_\mathrm{LS} \Gamma_\mathrm{SS} \mathbf{a}^\mathrm{true}_\mathrm{SS}\,,}\)}
\endgroup
\label{eq:shift_least_square}
\end{equation}
where~$\Gamma^+_\mathrm{LS} = \left( \Gamma^T_\mathrm{LS} \Gamma_\mathrm{LS} \right)^{-1} \Gamma^T_\mathrm{LS}$ is the Moore-Penrose inverse, and~$^T$ indicates the transpose (see also appendix~\ref{app:smallscale_leakage_complementary} for a derivation of this result).
The set of harmonic coefficients that minimize the~$\chi^2_\mathrm{LMS}$ statistics is not the set of ``true'' large-scale ones due to a contamination of small-scale modes.
This is what we refer to as ``small-scale leakage''.

\begin{figure}[ht]
    \centerline{
    \includegraphics[width=\textwidth]{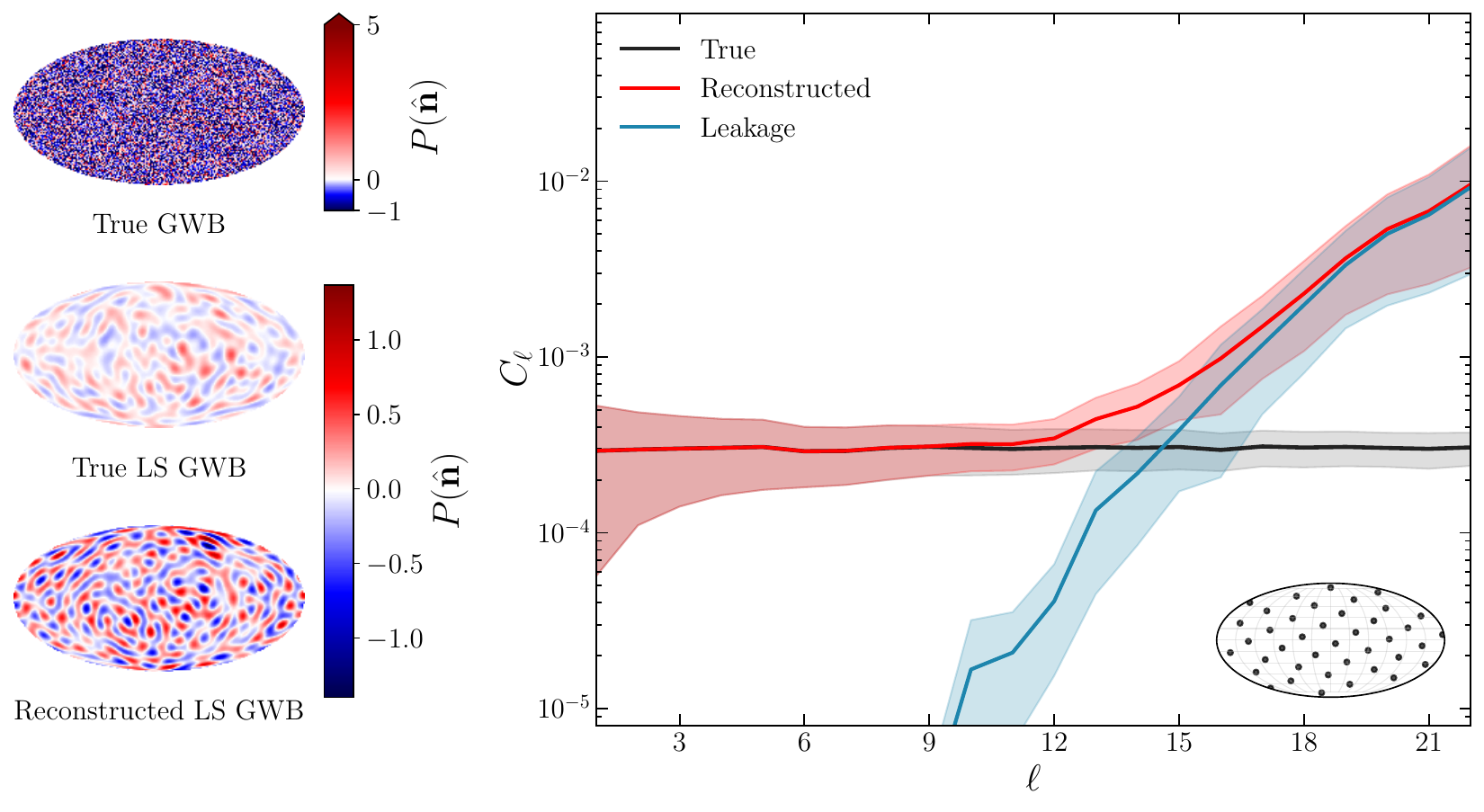}}
    \caption{\textit{Left panels}: individual realization of a true GWB with power up to~$\lmaxGWB=250$ (\textit{top panel}), its large-scale component with power up to~$\lmaxrec=22$ (\textit{central panel}), and reconstructed large-scale background, including small-scale leakage, up to~$\lmaxrec=22$ (\textit{bottom panel}). 
    \textit{Right panel}: envelopes of true (\textit{black}), reconstructed (\textit{red}), and leakage (\textit{blue}) correlated residual angular power spectra for~$10^3$ realizations of the GWB and the I34 pulsar geometry.
    Solid lines indicate the median value of angular power spectra.}
    \label{fig:SSL_pure_geometry}
\end{figure}

The implications of this fact can be immediately understood by constructing an estimator of the angular power spectrum.
Consider a vector~$\mathbf{a}^T_\ell = \left(a_{\ell\; -\ell}, \cdots, a_{\ell m}, \cdots , a_{\ell\; \ell}\right) = \mathbb{P}_\ell \mathbf{a}_\mathrm{LS}$ containing all the harmonic coefficients corresponding to a given multipole~$\ell$, where the explicit form of the projection matrix~$\mathbb{P}_\ell$ is reported in appendix~\ref{app:smallscale_leakage_complementary}.
For each multipole, the estimator of the angular power spectrum reads as
\begin{equation}
    \hat{C}_\ell = \frac{\mathbf{a}^T_\ell \mathbf{a}_\ell}{2\ell+1} = \frac{\left| \mathbb{P}_\ell \mathbf{a}^\mathrm{true}_\mathrm{LS} + \mathbb{P}_\ell \Gamma^+_\mathrm{LS} \Gamma_\mathrm{SS} \mathbf{a}^\mathrm{true}_\mathrm{SS} \right|^2}{2\ell+1}.
    \label{eq:cl_estimator}
\end{equation}
Therefore, as we show in appendix~\ref{app:smallscale_leakage_complementary}, the expectation value of this estimator is given by
\begin{equation}
    C^\mathrm{rec}_\ell = \left\langle \hat{C}_\ell \right\rangle = C^\mathrm{true}_\ell + C^\mathrm{leakage}_\ell,
    \label{eq:mode_mixing_pure}
\end{equation}
where~$C^\mathrm{rec}_\ell$ and~$C^\mathrm{true}_\ell$ are the reconstructed and true large-scale angular power spectrum, respectively.
Although at the level of the spherical harmonic coefficients this effect is referred to ``confusion noise'' in some portions of the literature, the effect of such a term at the level of the estimated angular power spectrum is that of a bias.
In other words, the leakage term should be interpreted as a realization-level contamination of the recovered large-scale coefficients and, after squaring and ensemble averaging, as a positive bias of the recovered angular power spectrum estimator.
This is directly analogous to the noise bias that has to be subtracted in CMB quadratic power spectrum estimators~\cite{Bond:1998zw}.

The leakage contribution explicitly reads as
\begin{equation}
\begingroup
  \setlength{\fboxrule}{1pt}% border thickness (default ~0.4pt)
  \setlength{\fboxsep}{6pt}% padding inside the box
  \scalebox{1.}{\(\boxed{C^\mathrm{leakage}_\ell = \frac{1}{2\ell+1} \left\langle \left|\mathbb{P}_\ell \Gamma^+_\mathrm{LS} \Gamma_\mathrm{SS} \mathbf{a}^\mathrm{true}_\mathrm{SS} \right|^2 \right\rangle = \sum_{\ell'=\lmaxrec+1}^{\lmaxGWB} M_{\ell \ell'} C^\mathrm{true}_{\ell'}\,,}\)}
\endgroup
\label{eq:cl_leakage_fitting}
\end{equation}
and~$M_{\ell \ell'}$ is a \textit{mode-mixing} function defined in appendix~\ref{app:smallscale_leakage_complementary} that depends exclusively on the geometry of the pulsar distribution.
This additional variance is sourced exclusively by the power at small scales, as the upper and lower bounds of the summation indicate.
Any term containing mixed large- and small-scale harmonic coefficients has a zero expectation value since the different multipoles are not correlated.

In figure~\ref{fig:SSL_pure_geometry} we compare the reconstructed (biased) angular power spectra with the true one for~$10^3$ realizations, and showcase that the additional variance is precisely described by the small-scale leakage formula derived in equation~\eqref{eq:cl_leakage_fitting}.
The relevance of this effect is displayed also on the left panels of the figure, where we show how the reconstructed map obtained with the harmonic coefficients of equation~\eqref{eq:shift_least_square} is significantly different from the original one, and it even reaches non physical values of~$P(\nhat) < -1$ because of the small-scale contamination.
The importance of this effect cannot be understated: due to the unmodeled small-scale power, the reconstructed large-scale angular power spectrum is overestimated by at least one order of magnitude with respect to its true value, far exceeding the level of uncertainty associated with cosmic variance~$\sigma_\mathrm{CV}/C_\ell = \sqrt{2/(2\ell+1)}$, which is of order~$20\%$ and~$15\%$ for~$\ell=22$ and~$\ell=46$, respectively.

\begin{figure}[ht]
    \centerline{
    \includegraphics[width=\columnwidth]{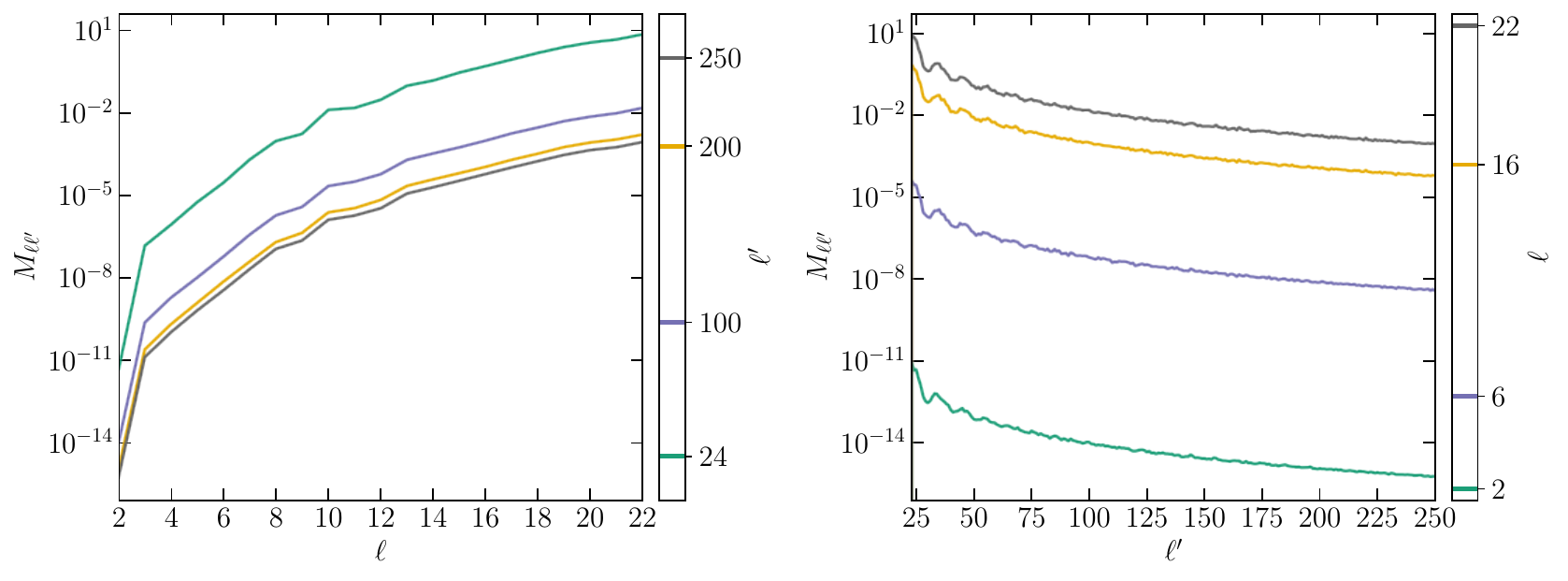}}
    \caption{Value of the mode-mixing function~$M_{\ell \ell'}$ from equation~\eqref{eq:cl_leakage_fitting} in terms of large-scale~$\ell$ and small-scale~$\ell'$ multipoles.
    This matrix quantifies how the unmodeled scales $\ell'$ bias the reconstructed $C_\ell$ (see equation~\eqref{eq:mode_mixing_pure}.
    In both panels we show how low large-scale multipoles are almost unaffected by small-scale leakage, however the situation radically changes for multipoles close to~$\lmaxrec$, where the order of magnitude of the mode-mixing function is around order unity.
    }
\label{fig:Mllprime_pure_geometry}
\end{figure}

The origin of this contamination can be traced back to the mode-mixing function~$M_{\ell \ell'}$ showed in figure~\ref{fig:Mllprime_pure_geometry} for the I34 pulsar geometry.
The structure of this function is such that large-scale multipoles far from~$\lmaxrec$ are minimally affected by small-scale leakage.
However, the magnitude of the mode-mixing function increases rapidly the closer we get to~$\lmaxrec$, reaching and exceeding values of order unity, driven mainly by small-scale multipoles close to~$\lmaxrec$.
This structure is consistent across different pulsar geometries, as we show in appendix~\ref{app:smallscale_leakage_complementary}.

%%%%%%%%%%%%%%%%%%%%%%%%%%%%%%%%%%%%%%%%%%%%%%%%%%%%%%%%%%%%%%%%%%%%%%%%%%%%%%%%%%%%%%%%%%%%%%%%%%%%%%%%%%%%%%%%%%%%%%%%%%%%%%%%%%%%

\subsection{Small-scale leakage - Statistical intuition}

The geometric picture presented above has, in reality, a deeper implication in terms of statistical inference of the harmonic coefficients, as in a likelihood-based analysis.
Harmonic coefficients in the weak field regime have a Gaussian distribution; therefore, we can introduce a Gaussian likelihood statistics as~\cite{alihaimoud2020}
\begin{equation}
    -2\log\mathcal{L}(\mathbf{a}_\mathrm{LS} | \boldsymbol{\RR}) = \left( \boldsymbol{\RR} - \boldsymbol{\RR}^\mathrm{th} \right)^T \mathrm{Cov}^{-1} \left( \boldsymbol{\RR} - \boldsymbol{\RR}^\mathrm{th} \right),
\label{eq:likelihood}
\end{equation}
where~$\boldsymbol{\RR}^\mathrm{th}$ is the vector of theoretical correlated residuals, and~$\mathrm{Cov} = \left \langle \left( \boldsymbol{\RR} - \boldsymbol{\RR}^\mathrm{th} \right) \left( \boldsymbol{\RR} - \boldsymbol{\RR}^\mathrm{th} \right)^T \right \rangle$ is a covariance matrix.\footnote{
The likelihood in equation~\eqref{eq:likelihood} is used exclusively to determine the harmonic coefficients that best match the realization of the Universe in which we live in.
It does not serve the purpose of estimating the posterior of the theoretical model parameters used to generate our set of realizations.
Therefore, the extra term~``$\log\mathrm{det} \mathrm{Cov}$'' that should appear in the case of Gaussian likelihoods has been removed from equation~\eqref{eq:likelihood} since it is just a fixed additive constant for the purpose at hand.
On the other hand, in a band-power estimation framework, where the theoretical~$C_\ell$ enter the covariance, this term is essential for regularization and must be retained.}
In the pure signal limit, as shown in appendix~\ref{app:smallscale_leakage_complementary}, the elements of the covariance matrix read as
\begin{equation}
    \mathrm{Cov}_{i_{pq} j_{rs}} = \sum_{\ell m} \left( \Gamma^{pr}_{\ell m} \Gamma^{qs}_{\ell m} + \Gamma^{ps}_{\ell m} \Gamma^{qr}_{\ell m} \right) \left(C_\ell + N^\mathrm{shot}_\ell \right),
    \label{eq:cov_puresignal}
\end{equation}
where~$N^\mathrm{shot}_\ell$ is the Poissonian shot-noise term due to the discreteness of the SMBHBs, and with the composite indices~$i_{pq}$ we explicitly indicate which pulsar pair corresponds to each row/column of the covariance matrix.

As we demonstrate in appendix~\ref{app:smallscale_leakage_complementary}, under the incorrect assumption that correlated residuals are described only by large-scale modes, i.e., that~$\boldsymbol{\RR}^\mathrm{th} = \Gamma_\mathrm{LS} \mathbf{a}_\mathrm{LS}$, we observe a shift in the maximum likelihood position given by
\begin{equation}
\mathbf{a}_\mathrm{LS} = \mathbf{a}_\mathrm{LS}^\mathrm{true} + \left(\Gamma^T_\mathrm{LS} \mathrm{Cov}^{-1} \Gamma_\mathrm{LS} \right)^{-1} \Gamma^T_\mathrm{LS} \mathrm{Cov}^{-1} \Gamma_\mathrm{SS} \mathbf{a}_\mathrm{SS}^\mathrm{true}\,.
\label{eq:shift_maximum_likelihood}
\end{equation}
Therefore, also in this formulation of the issue, the expectation value of the reconstructed harmonic coefficients contains a leakage term that reads as
\begin{equation}
   C^\mathrm{leakage}_\ell = \frac{\left\langle \left| \mathbb{P}_\ell \left(\Gamma^T_\mathrm{LS} \mathrm{Cov}^{-1} \Gamma_\mathrm{LS} \right)^{-1} \Gamma^T_\mathrm{LS} \mathrm{Cov}^{-1} \Gamma_\mathrm{SS} \mathbf{a}^\mathrm{true}_\mathrm{SS} \right|^2 \right\rangle}{2\ell+1} = \sum_{\ell'=\lmaxrec+1}^{\lmaxGWB} M'_{\ell \ell'} C^\mathrm{true}_{\ell'}\,.
\label{eq:cl_leakage_likelihood}
\end{equation}

Since the form and structure of the mode-mixing function~$M'_{\ell \ell'}$ is similar to that presented in figure~\ref{fig:Mllprime_pure_geometry}, we choose to report them both in appendix~\ref{app:smallscale_leakage_complementary}.
Moreover, since the case presented in section~\ref{subsec:smallscaleleakage_geometricalintuition} can be exactly recovered by assuming that~$\mathrm{Cov}=I_{N_\mathrm{pairs}}$, in the following we focus only on this interpretation, drawing a connection to the simple map-making process only when needed.

%%%%%%%%%%%%%%%%%%%%%%%%%%%%%%%%%%%%%%%%%%%%%%%%%%%%%%%%%%%%%%%%%%%%%%%%%%%%%%%%%%%%%%%%%%%%%%%%%%%%%%%%%%%%%%%%%%%%%%%%%%%%%%%%%%%%

\subsection{Regularization scheme and interplay with small-scale leakage}
\label{subsec:regularization_scheme}

Although the analysis presented above is statistically well posed, in reality its numerical structure is ill-conditioned.
In particular, we can foresee two potential sources of numerical instability in the likelihood analysis which might affect the inference of the harmonic coefficients.

The first source of instability comes from the likelihood itself.
Its origin can be traced back to the presence of quasi-singular pulsar configurations that make the~$\Gamma_\mathrm{LS}$ matrix lines almost linearly dependent.
In this instance, the~$\mathbf{a}_\mathrm{LS}$ space presents an almost-degenerate direction.
In this analysis, we introduce a \textit{ridge regularization} scheme to address this potentially critical issue.
We add a penalization term to equation~\eqref{eq:likelihood} given by~$\lambda \mathbf{a}^T_\mathrm{LS} \mathcal{P} \mathbf{a}_\mathrm{LS}$, where~$\lambda > 0$ is the regularization parameter, and~$\mathcal{P}$ is a symmetric, positive semi-definite penalty matrix.
The new maximum likelihood position is now given by
\begin{equation}
    \mathbf{a}_\mathrm{LS} = \left[ I - \left(\Gamma^T_\mathrm{LS} \mathrm{Cov}^{-1} \Gamma_\mathrm{LS} + \lambda\mathcal{P} \right)^{-1} \lambda\mathcal{P} \right] \mathbf{a}^\mathrm{true}_\mathrm{LS} + \left(\Gamma^T_\mathrm{LS} \mathrm{Cov}^{-1} \Gamma_\mathrm{LS} + \lambda\mathcal{P} \right)^{-1} \Gamma^T_\mathrm{LS} \mathrm{Cov}^{-1} \Gamma_\mathrm{SS} \mathbf{a}^\mathrm{true}_\mathrm{SS},
\label{eq:maximumlikelihoodshift_regularization}
\end{equation}
as we show in appendix~\ref{app:regularization}.
First of all, we note that regularization schemes come with an intrinsic trade-off: even though they remove almost-degenerate directions from the likelihood analysis, they are also responsible for the introduction of an unavoidable bias in the determination of the true large-scale coefficients, as we read from the first term of equation~\eqref{eq:maximumlikelihoodshift_regularization}.
Second, we observe that the leakage term still appears in the second term of equation~\eqref{eq:maximumlikelihoodshift_regularization}; however, we can already foresee that this term is suppressed for large values of the regularization parameter.
Although this choice of regularization scheme is arbitrary, the point we are raising is not: removing numerical instabilities from the analysis in an uncontrolled fashion has the potential of biasing the inference of an unknown physical signal.

The second potential source of instability is due to an ill-conditioned covariance matrix.
Also in this case, in the presence of pulsar geometries that present near-perfect degeneracies, the inverse of the covariance matrix could be numerically unstable.
A typical solution involves the introduction of a regularized version of the covariance matrix,~$\mathrm{Cov}^{-1}_\mathrm{reg}$, see, e.g., appendix~\ref{app:regularization} for the practical implementation used in this work.
In terms of the inferred contamination in the value of large-scale coefficients, this regularization does not solve the small-scale leakage, since this only exchanges~$\mathrm{Cov}^{-1}$ for~$\mathrm{Cov}^{-1}_\mathrm{reg}$ in equation~\eqref{eq:maximumlikelihoodshift_regularization}.
The large-scale coefficients are still contaminated by a small-scale leakage, even though by a different amount.
We note that it is not possible to disentangle the small-scale leakage and the additional effect due to covariance regularization.

Once both of these regularization schemes are taken into account, we find that the reconstructed angular power spectrum reads as
\begin{equation}
    C^\mathrm{rec}_\ell = C^\mathrm{true}_\ell - C_\ell^\mathrm{reg} + C^\mathrm{leakage}_\ell,
\end{equation}
where the \textit{regularization bias} term~$C^\mathrm{reg}_\ell$ is given by
\begin{equation}
    C^\mathrm{reg}_\ell = \sum^{\lmaxrec}_{\ell'=2} K_{\ell \ell'} C^\mathrm{true}_{\ell'},
    \label{eq:reg_bias_cell}
\end{equation}
and~$K_{\ell \ell'}$ is a second, regularization scheme dependent, mode-mixing matrix, this time coupling different large-scale modes.
Its explicit form is reported in appendix~\ref{app:regularization}, alongside with the form of the new mode-mixing matrix~$M''_{\ell \ell'}$ appearing in the leakage term
\begin{equation}
    C^\mathrm{leakage}_\ell = \sum_{\ell'=\lmaxrec+1}^{\lmaxGWB} M''_{\ell \ell'} C^\mathrm{true}_{\ell'}.
    \label{eq:Msec_cell}
\end{equation}
The functional form of~$M''_{\ell \ell'}$ and~$K_{\ell \ell'}$ is also shown in appendix~\ref{app:regularization}.

\begin{figure}[ht]
    \centerline{
    \includegraphics[width=\columnwidth]{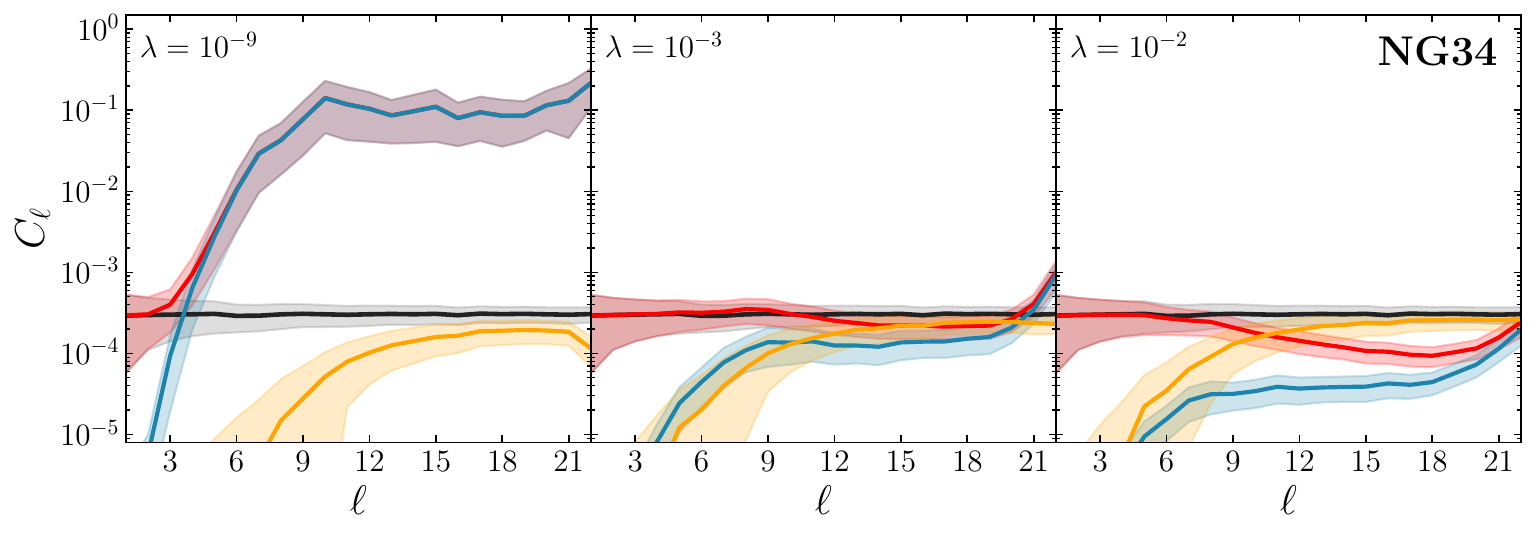}}
    \centerline{
    \includegraphics[width=\columnwidth]{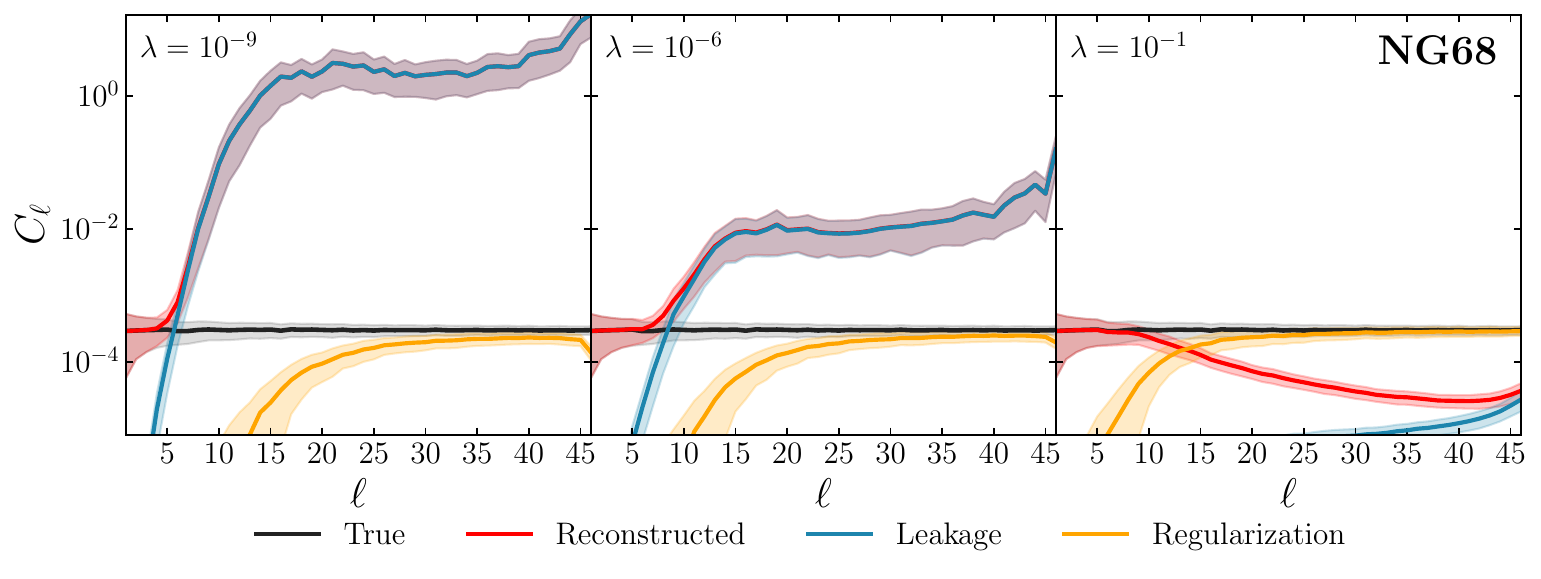}}
    \caption{Envelopes of true (\textit{black}), reconstructed (\textit{red}), leakage (\textit{blue}) and regularization (\textit{orange}) $C_\ell$ for~$10^3$ realizations of the GWB and the NG34 (\textit{top panels}) and NG68 (\textit{bottom panels}) pulsar geometry.
    Different panels show the effect of different values of the ridge regularization parameter, indicated on the top left corner.
    Solid lines indicate the median value of angular power spectrum. 
    Although a mild regularization appears to successfully suppress the small-scale leakage effect, more aggressive choices suppress the reconstructed signal below its true value. 
    Since the true value of GWB angular power spectrum is not known a priori, it is not possible to fine-tune the regularization process to filter out the spurious effect.}
\label{fig:regularization_leakage_interplay}
\end{figure}

We show in figure~\ref{fig:regularization_leakage_interplay} how different choices of the ridge parameter affect the reconstruction of the angular power spectra for NG34 and NG68 pulsar configurations in the top and bottom panels, respectively.
First, we observe how for small values of~$\lambda$ the amount of small-scale leakage extends to very low multipoles for both configurations, especially compared to the isotropic case of section~\ref{subsec:smallscaleleakage_geometricalintuition}.
Increasing the magnitude of~$\lambda$ effectively reduces the leakage contribution; however, it also increases the regularization bias, making it of the same order of magnitude of the true angular power spectrum, thus effectively suppressing the reconstructed signal.
This phenomenology is supported by the overall form of the~$M''_{\ell \ell'}$ and~$K_{\ell\ell'}$ functions, as we show in appendix~\ref{app:regularization}.
In particular, we notice~$M''_{\ell \ell'}$ presents the same shape as in the purely geometric case, but its amplitude decreases as the ridge parameter increases, i.e., $M''_{\ell \ell'}(\lambda \to \infty)\to 0$, effectively canceling the leakage term.
In contrast, we observe for the regularization matrix that~$K_{\ell \ell'}(\lambda\to\infty) \to \delta_{\ell\ell'}$, so that~$C^\mathrm{reg}_\ell \to C^\mathrm{true}_\ell$ and a null signal is reconstructed.
The phenomenology in the isotropic and uniformly distributed pulsar is qualitatively identical to the NANOGrav pulsar geometry, therefore it is not displayed.

Finally, it is natural to wonder whether small-scale leakage can be eliminated by choosing a larger reconstruction multipole~$\ell^\mathrm{rec}_\mathrm{max}$ or making an appropriate choice of the ridge parameter.
Regarding the former, we note that increasing the dimension of the recovery parameter space could be helpful in reducing the small-scale contamination caused by a conservative truncation of the signal model.
However, a larger recovery space does not eliminate unresolved-mode contamination by itself: introducing additional parameters beyond the number of independent constraints from pulsar pairs necessarily introduces additional degeneracies.
Furthermore, directions in parameter space that are poorly constrained by the array geometry are not guaranteed at all to be aligned with directions that are contaminated by leakage; hence, direct regularization of the Fisher matrix does not guarantee that the leakage is removed along with the poorly constrained modes.
For what concerns the choice of ridge parameter, unfortunately, no choice can be made a priori, since we lack any knowledge regarding the amount of power at small scales.
In other words, since we are not aware from first principles of the magnitude of the small-scale leakage, we cannot use this regularization scheme to recover the true power spectrum in the multipole range that has already been contaminated by small-scale power.

%%%%%%%%%%%%%%%%%%%%%%%%%%%%%%%%%%%%%%%%%%%%%%%%%%%%%%%%%%%%%%%%%%%%%%%%%%%%%%%%%%%%%%%%%%%%%%%%%%%%%%%%%%%%%%%%%%%%%%%%%%%%%%%%%%%%

\section{Impact of pulsar timing noise}
\label{sec:pulsar_noise}

Although so far we considered a purely deterministic analysis in which pulsar noise plays no role, this is not the case in a real-life scenario.
Here, we want to demonstrate that the small-scale leakage exists even in a scenario where noise is included, even though some simplifying assumptions are still taken to maintain the formalism completely analytical.
In this case, the inclusion of the noise transforms the form of the full covariance matrix introduced in the likelihood, which now receives an additional contribution and reads as
\begin{equation}
    \mathrm{Cov}_{i_{pq} j_{rs}} = \left[ \delta^K_{pr} \delta^K_{qs} + \delta^K_{ps} \delta^K_{qr} \right] \sigma^2_p \sigma^2_q + \sum_{\ell m} \left( \Gamma^{pr}_{\ell m} \Gamma^{qs}_{\ell m} + \Gamma^{ps}_{\ell m} \Gamma^{qr}_{\ell m} \right) \left(C_\ell + N^\mathrm{shot}_\ell \right).
\label{eq:full_cov}
\end{equation}

\begin{figure}[ht]
    \centerline{
    \includegraphics[width=\columnwidth]{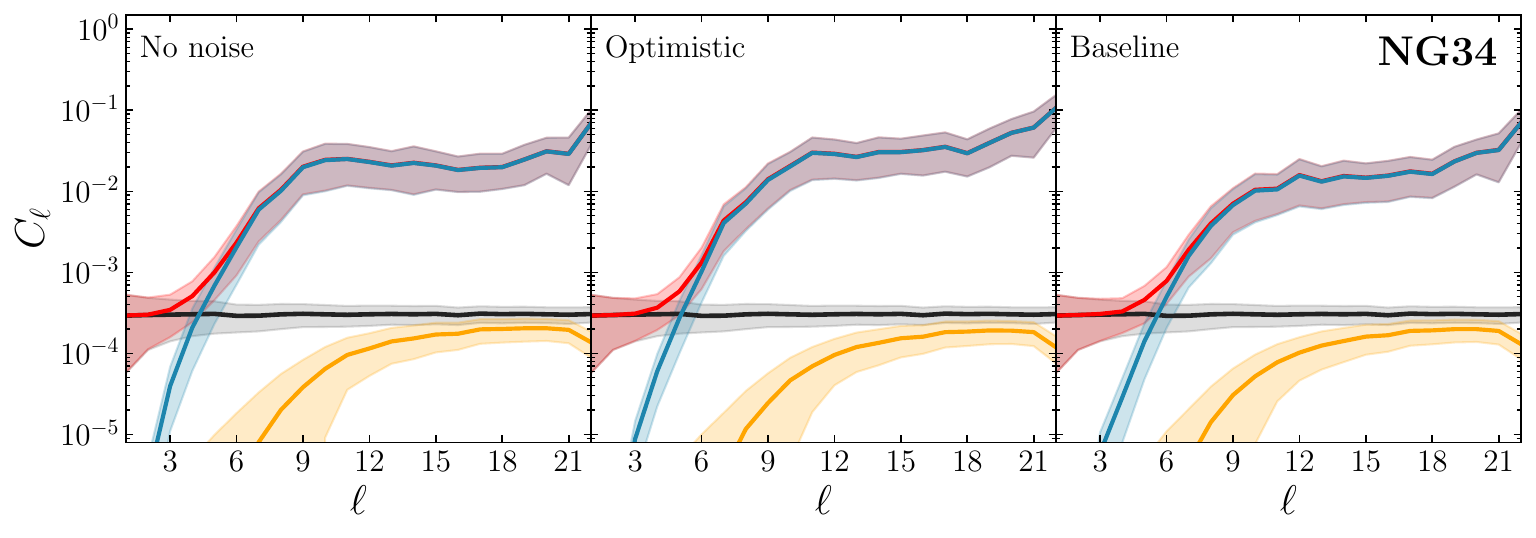}}
    \centerline{
    \includegraphics[width=\columnwidth]{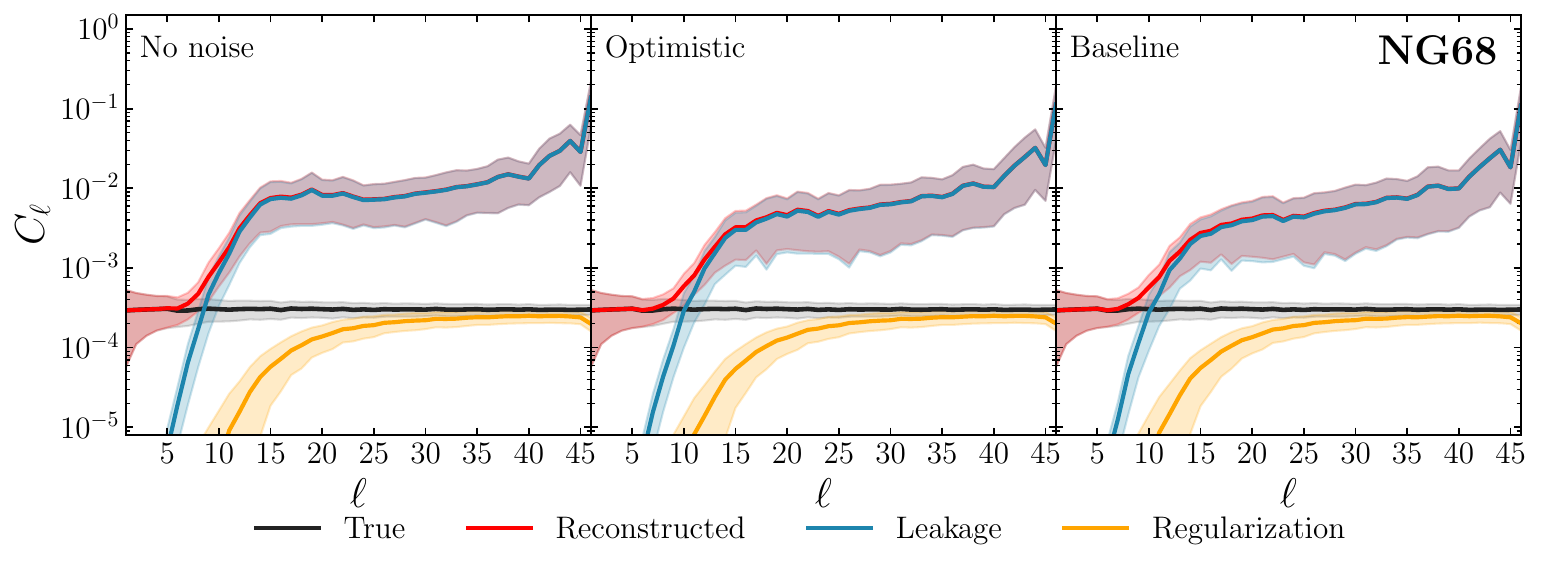}}
    \caption{Same of figure~\ref{fig:regularization_leakage_interplay}, but for different amounts of injected pulsar red noise. 
    Although in both cases the injected noise does not significantly affect the reconstruction of the signal, we observe as it increases the variance of the bands.}
\label{fig:regularization_leakage_noise_interplay}
\end{figure}

We created a realistic set of pulsar noise values by rescaling the current noise budget of pulsars in the NANOGrav 15-year dataset in the~$f=4\ \mathrm{nHz}$ frequency bin~\cite{nanograv_detector}.
The magnitude of this rescaling makes the error budget on average compatible with the detection of GWB anisotropies.
In particular, we define a \textit{baseline} case, where the noise and the signal terms appearing in equation~\eqref{eq:full_cov} have approximately the same order of magnitude (i.e.,~$\sim 50\%$ of the pulsar pairs are signal dominated); and an \textit{optimistic} case, where the noise is on average subdominant with respect to the signal (i.e.,~$\sim 90\%$ of the pulsar pairs are signal dominated).
We show in figure~\ref{fig:regularization_leakage_noise_interplay} the impact of considering noise in the measurements of the correlated residuals.
As we observe by contrasting figures~\ref{fig:regularization_leakage_interplay} and~\ref{fig:regularization_leakage_noise_interplay}, the inclusion of this additional term in the covariance does not substantially alter the phenomenology already described in the previous section.
This is not unexpected, since also in this case the position of the maximum likelihood has still the form reported in equation~\eqref{eq:maximumlikelihoodshift_regularization}.
Moreover, for any practical purpose, one could even think of the noise contribution as some form of natural regularization of the covariance matrix, which in this case necessitates a milder regularization cut-off.
However, the effect of small-scale leakage is still present in all its relevance, regardless of the noise configuration, the pulsar geometry, or the regularization scheme (even if, for conciseness, we show only one case for the latter).
Also in this scenario, the phenomenology in the isotropic and uniformly distributed pulsar is qualitatively identical to the NANOGrav pulsar geometry, therefore it is not displayed.

%%%%%%%%%%%%%%%%%%%%%%%%%%%%%%%%%%%%%%%%%%%%%%%%%%%%%%%%%%%%%%%%%%%%%%%%%%%%%%%%%%%%%%%%%%%%%%%%%%%%%%%%%%%%%%%%%%%%%%%%%%%%%%%%%%%%

\section{Discussion and conclusions}
\label{sec:discussion_conclusions}

Since each pulsar timing residual measurement is sensitive to the GW power emitted across the entire sky, unresolved small-scale power unavoidably leaks into the large-scale one when uncontrolled partial reconstruction of the GWB is attempted.
This effect is particularly significant for the case in which the origin of the background is astrophysical, since the GW power generated by SMBHBs tracing the cosmic web is spread across an extremely wide range of angular scales. 
The unmodeled power beyond the current resolution limits of~$\lmaxrec \sim 10$ does not simply vanish from the correlated timing residuals; therefore, it has the potential to be misinterpreted as large-scale power.
Our conclusions broadly apply to the methodology commonly used for PTA anisotropy analyses, such as finite-resolution reconstructions based on truncated spherical-harmonic or pixel expansions~\cite{agazie2023_anisotropies, Taylor:2015udp, Miles_2024}.

The magnitude of this systematic effect strongly depends on the amount of power at small scales.
In this work, we show that, for realistic realizations of the astrophysical background, this contamination exceeds the cosmic variance level by at least one order of magnitude across a broad range of scales, fundamentally altering how we interpret anisotropy measurements from PTA experiments.
These numbers should be interpreted as conservative, as we artificially removed any power existing at scales~$\ell > \ell_{\max}^{\rm GWB} = 250$ to reduce computational costs.

Furthermore, as demonstrated in this article, the presence of this systematic error is independent of the specific details of the pulsar spatial configuration, the regularization scheme introduced in the analysis, and the presence of pulsar noise.
The practical severity of the effect is set by the interplay between the true small-scale angular power spectrum and the mode-coupling efficiency associated with each PTA pulsar geometry.
The contamination of very small scales may or may not be suppressed by the PTA response, but unresolved modes near the effective resolution scale can still contaminate the recovered low-multipole sector if they are not modeled or marginalized over.
Realistic noise budgets (compatible with anisotropy detection) do not eliminate leakage: noise changes the covariance that enters the mode-mixing kernel, but the systematic contamination from unmodeled small-scale power persists.
Moreover, the regime in which anisotropy detection becomes plausible is precisely the regime in which a significant fraction of pulsar pairs is signal-dominated and the leakage is largest.
Finally, we show how regularization represents both a necessity and a compromise, since it introduces its own regularization bias in the reconstructed angular power spectra, and therefore it cannot be thought of as a possible solution to the small-scale leakage issue.

Perfect reconstruction of the large-scale angular structure of the background appears challenging without any prior knowledge of the amount of small-scale power, since it would require the fulfillment of at least one of these conditions: \textit{(i)} complete absence of small-scale power, i.e.,~$\mathbf{a}_\mathrm{SS} = \mathbf{0}$; or \textit{(ii)} an astonishing large resolution, so that $\RR_{\rm SS}\ll\RR_{\rm LS}$ for all pairs.
Since none of these conditions can be achieved for realistic PTA experiments, the bias introduced by the small-scale leakage effect is unavoidable, and some mitigation strategy should be implemented in the statistical analysis.

A possible way forward in solving this issue would be to create an estimator that has some intrinsic form of ``orthogonality'' between large- and small-scale correlated residuals.
An alternative way forward could be represented by modeling and marginalizing over concrete scenarios with a characteristic amount of small-scale power.
In this case, we should talk about \textit{conditional reconstruction} of the GWB anisotropy, and the scope of the reconstruction will be limited by the intrinsic robustness of the assumption about small-scale power.
In other words, if a wrong small-scale model or prior are chosen, the reconstructed large-scale power will still be biased following the same logic of this work.
Finally, we stress that, although we work in a spherical-harmonic basis for the sake of concreteness, the existence of this underlying issue is not restricted to this basis.
Equivalent statements hold in a pixel basis or in a principal-map basis: whenever the data are interpreted in a truncated reconstruction space while the true signal contains additional angular structure, the unresolved sector contaminates the recovered large-scale modes unless explicitly modeled or marginalized over.

Additionally, the presence of this systematic effect should also be addressed on the side of numerical simulations.
If the simulated signal does not incorporate a realistic realization of the power present at small scales in realistic, physically motivated scenarios, the subsequent analysis will be automatically blind to the presence of this effect.
Moreover, since such simulations are also used to validate optimal estimators, the statistical pipeline developed to analyze current and future datasets might not reach the desired level of accuracy.

On the physical interpretation side, the consequences of small-scale leakage are potentially catastrophic.
The distorted maps and angular power spectra are fundamentally incompatible with any large-scale structure pattern, and thus with any astrophysical interpretation.
Therefore, any sort of ``cross-correlation'' analysis with galaxy catalogs, either to locate sources in host galaxies or to study the clustering properties of SMBHBs, will return spurious correlations sourced by this systematic effect.

Although this work considers only the weak-field regime scenario for GWB anisotropies, we do not expect the situation to improve once individual sources are introduced into the analysis.
In contrast, loud continuous-wave sources create localized hot spots and are associated with scale-invariant angular power spectrum with power spread over a wide range of scales.
Moreover, their presence also breaks the assumption of Gaussianity of the inferred harmonic coefficients. 
Therefore, in addition to the guaranteed presence of a small-scale leakage, we expect further complications to arise in terms of dealing with the statistics of the correlated residuals.
In this sense, the nature of the leakage effect is fully general in different scenarios.

Looking forward, CMB-inspired solutions, such as designing unbiased band-power estimation using a quadratic estimator~\cite{Seljak:1997wx, Bond:1998zw} might be proven useful for the PTA purposes.
This sort of estimators would include a small number of well-resolved angular bins together with one or, more likely, many nuisance bins absorbing power from the unresolved sector, with explicit noise-bias subtraction.
The leakage contribution~$C_\ell^{\rm leakage}$ derived in this work is precisely the quantity that such an estimator would need to model or marginalize.
Developing this framework for PTA data is part of our planned follow-up work.

Forward modeling and trans-dimensional approaches have also been proposed as promising directions for PTA data analysis.
However, no existing work has demonstrated that these methods eliminate the contamination from unresolved small-scale power.
The physical effect that correlated timing residuals account for GW power distributed over all angular scales persists regardless of the inference framework.
Forward modeling requires prior assumptions on the small-scale power distribution, which depends sensitively on the poorly constrained properties of the SMBHB population.
Trans-dimensional approaches let the data select model complexity, but the finite number of pulsar pairs still limits the number of independent constraints; when the sampler selects more parameters than well-constrained modes, the posterior is dominated by the prior, effectively implementing a form of regularization with its associated bias--variance tradeoff.

Statistical uncertainties will continue to decrease as PTAs start monitoring a larger number of pulsars and extend their observation baselines.
However, systematic effects will not disappear and will eventually dominate the total error budget, unless addressed explicitly. 
In particular, small-scale leakage represents a fundamental limitation of current analysis methods when reconstructing the GW sky from PTA data; thus, it can be mitigated but not fully eliminated by increasing the number of pulsars. 
In this sense, understanding and mitigating these systematic effects will mark a crucial step toward reliable characterization of the nHz GWB angular structure. 

%%%%%%%%%%%%%%%%%%%%%%%%%%%%%%%%%%%%%%%%%%%%%%%%%%%%%%%%%%%%%%%%%%%%%%%%%%%%%%%%%%%%%%%%%%%%%%%%%%%%%%%%%%%%%%%%%%%%%%%%%%%%%%%%%%%%

\section*{Acknowledgements}

The authors thank Deepali Agarwal, Mesut \c{C}al{\i}\c{s}kan, Neha Anil Kumar, Michele Liguori, Giorgio Mentasti, Nihan Pol, Luca Reali, Joseph Romano, Stephen Taylor, and Benjamin Wandelt for useful discussions.
The authors also thank Bjorn Larsen for providing the red noise data for the NANOGrav set of pulsars.

This work is partly supported by ICSC - Centro Nazionale di Ricerca in High Performance Computing, Big Data and Quantum Computing, funded by European Union - NextGenerationEU.
NB acknowledges partial support by PRD/ARPE 2022 ``Cosmology with Gravitational waves and Large Scale Structure - CosmoGraLSS'' and by the European Union's Horizon Europe research and innovation program under the Marie Sk\l{}odowska-Curie grant agreement no. 101207487 (GWSKY - Mapping the Universe with Gravitational Waves).
CMFM acknowledges support from the National Science Foundation from Grants NSF PHY-2020265 and AST-2414468. This work was also supported by the Flatiron Institute, part of the Simons Foundation.
AR acknowledges funding from the Italian Ministry of University and Research (MIUR) through the ``Dipartimenti di eccellenza'' project ``Science of the Universe''.

%%%%%%%%%%%%%%%%%%%%%%%%%%%%%%%%%%%%%%%%%%%%%%%%%%%%%%%%%%%%%%%%%%%%%%%%%%%%%%%%%%%%%%%%%%%%%%%%%%%%%%%%%%%%%%%%%%%%%%%%%%%%%%%%%%%%

\appendix
\section{Persistence of small-scale residuals}
\label{app:persistence_smallscale_residuals}

\begin{figure}[ht]
    \centerline{
    \includegraphics[width=\columnwidth]{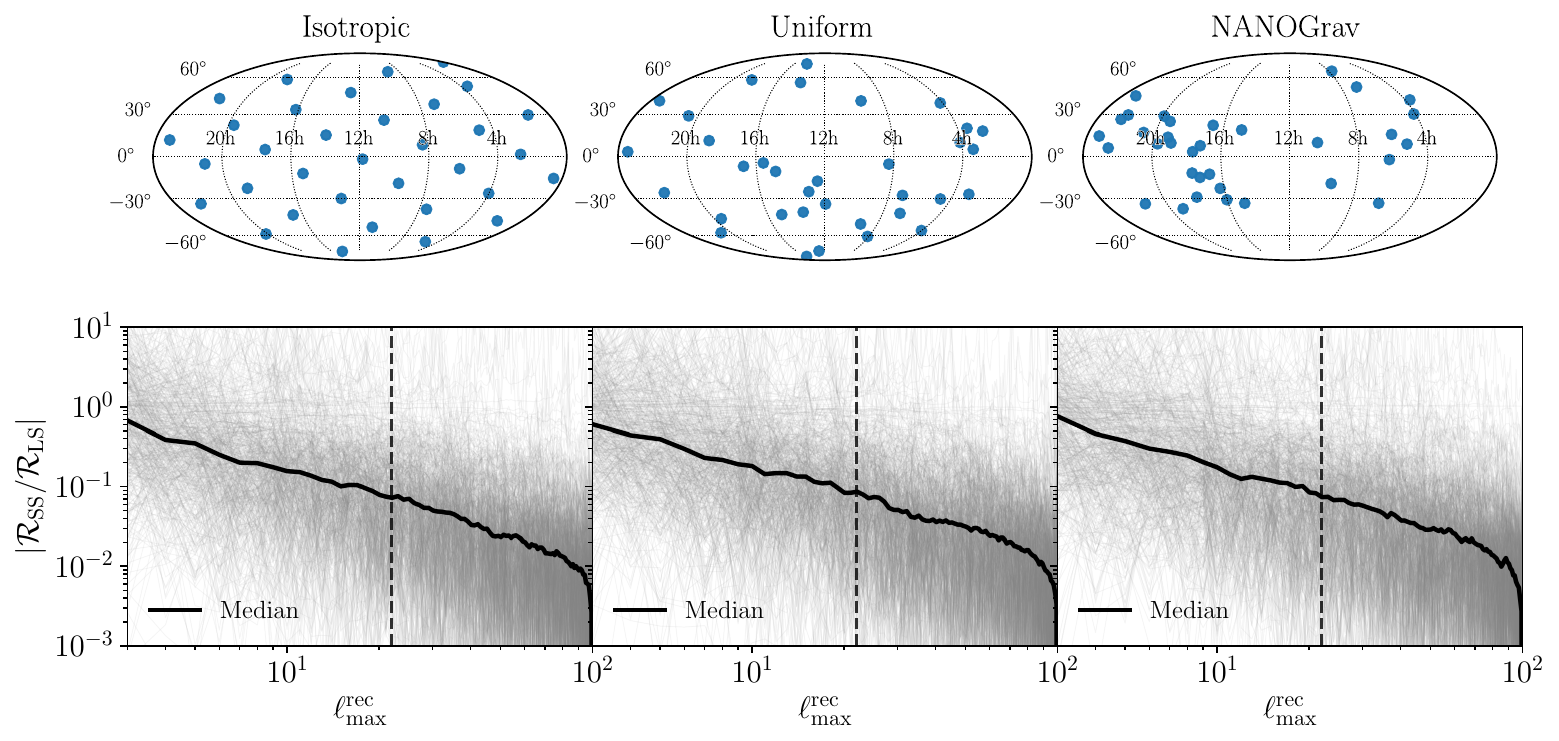}}
    \caption{\textit{Top panels}: isotropic (I34), uniform (U34) and NANOGrav (NG34) samples of pulsars considered in this work. 
    \textit{Bottom panels}: absolute value of the ratio of small-scale to large-scale correlated residuals for all the pulsar pairs in the sample.
    The \textit{black solid line} represents the median value of~$|\mathcal{R}_\mathrm{SS}/\mathcal{R}_\mathrm{LS}|$.}
\label{fig:RSS_over_RLS_general}
\end{figure}

\begin{figure}[ht]
    \centerline{
    \includegraphics[width=\columnwidth]{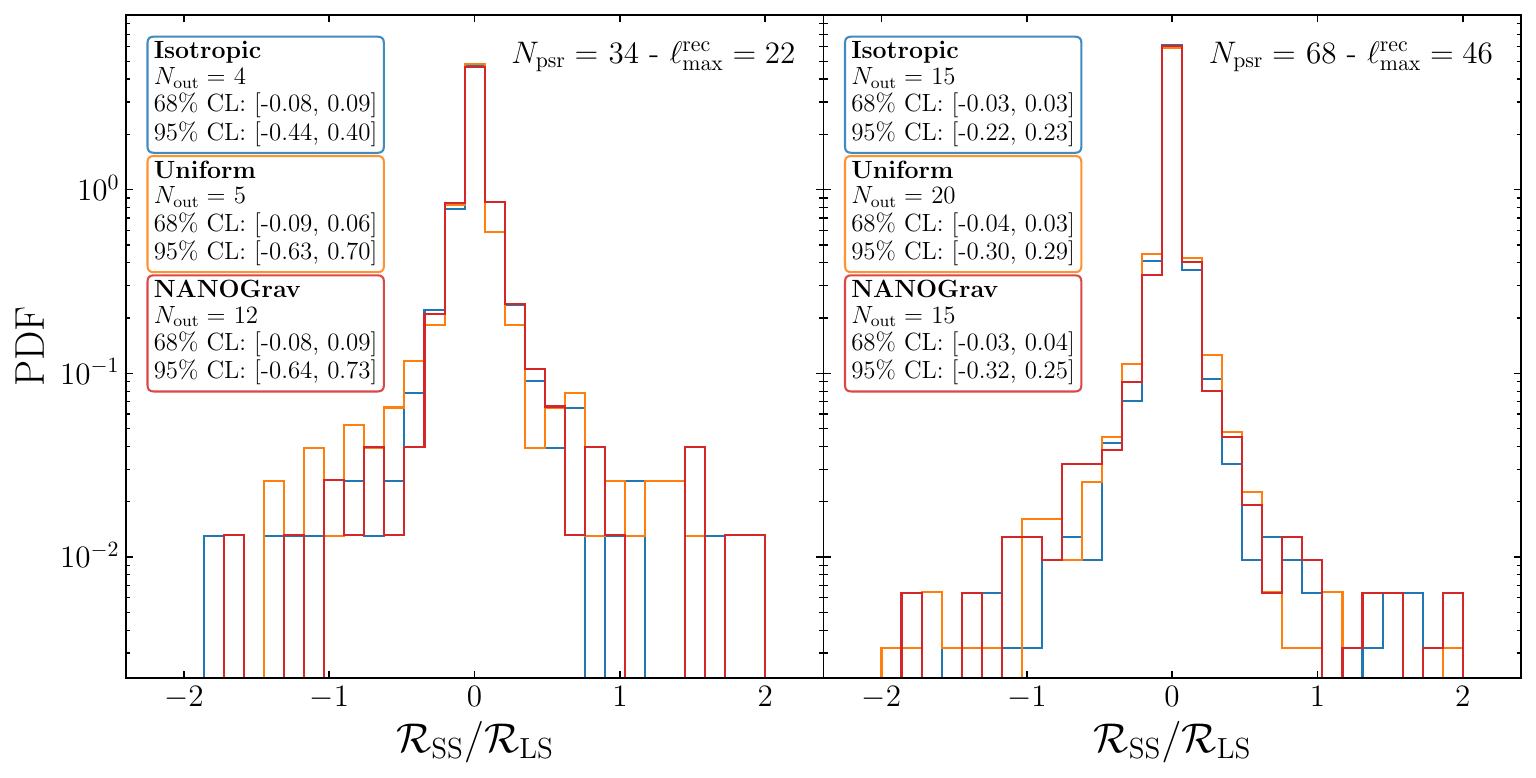}}
    \caption{Probability distribution function (PDF) of the $\RR_\mathrm{SS}/\RR_\mathrm{LS}$ correlated residual ratio for the $N_\mathrm{psr}=34$ (\textit{left panel}) and~$N_\mathrm{psr}=68$ (\textit{right panel}) pulsar configurations.
    Each PDF is computed at the maximum reconstruction multipole indicate in the top right corner.
    \textit{Blue}, \textit{orange}, and \textit{red} lines indicate the isotropic, uniform, and NANOGrav configurations, respectively.
    The tables on the top left corners report for each distribution the number of outliers~$N_\mathrm{out}$ such that~$|\RR_\mathrm{SS}/\RR_\mathrm{LS}|>2$, and the~$68\%,95\%$ confidence limits (CL).
    As the confidence limits suggest, even though the PDFs seem Gaussian, they have generally larger tails.}
\label{fig:RSS_over_RLS_pdf}
\end{figure}

The existence of small-scale correlated residuals is a physical effect completely independent of the geometrical distribution of pulsars.
We show in figure~\ref{fig:RSS_over_RLS_general} the absolute value of the small-scale to large-scale correlated residuals ratio for all possible pairs of the I34, U34 and NG34 geometrical configurations of pulsars.
As it immediately appears, especially for very small maximum reconstruction multipoles, this ratio is order unity.
Therefore, by attempting any partial reconstruction of a GWB anisotropic map, we are effectively misplacing almost the entire amount of small-scale power into large-scale modes.
This behavior is consistent across different geometries, as it becomes even more evident by comparing the probability distribution functions of the small- to large-scale correlated residual ratio in figure~\ref{fig:RSS_over_RLS_pdf} for the pulsar geometries considered in this work at their respective maximum reconstruction multipole.
We also report in the tables in the top left corners of the panels some statistical quantities of interest that describe such probability distribution functions.
We observe the existence of a small, but not zero, number of ``outlier'' pulsar pairs~$N_\mathrm{out}$, for which this ratio takes values~$\left| \RR_\mathrm{SS}/\RR_\mathrm{LS} \right| > 2$.
Moreover, by comparing the proportion between the~$68\%$ and~$95\%$ CL regions, we note that these distributions have strong non-Gaussian tails, allowing for a large number of pairs where the error on the assignment of power is almost order unity.
The dispersion of the probability distribution function at multipole~$\ell<\lmaxrec$ is larger than in the case shown in the figure.

%%%%%%%%%%%%%%%%%%%%%%%%%%%%%%%%%%%%%%%%%%%%%%%%%%%%%%%%%%%%%%%%%%%%%%%%%%%%%%%%%%%%%%%%%%%%%%%%%%%%%%%%%%%%%%%%%%%%%%%%%%%%%%%%%%%%%%%%%%

\section{Complementary material on small-scale leakage derivation}
\label{app:smallscale_leakage_complementary}

%%%%%%%%%%%%%%%%%%%%%%%%%%%%%%%%%%%%%%%%%%%%%%%%%%%%%%%%%%%%%%%%%%%%%%%%%%%%%%%%%%%%%%%%%%%%%%%%%%%%%%%%%%%%%%%%%%%%%%%%%%%%%%%%%%%%%%%%%%

\subsection{Geometrical intuition}

The minimum of the~$\chi^2_\mathrm{LMS}$ is found by solving the \textit{normal equations}
\begin{equation}
    \nabla_{\mathbf{a}_\mathrm{LS}} \chi^2_\mathrm{LMS} \simeq \Gamma^T_\mathrm{LS} \left( \Gamma_\mathrm{LS} \mathbf{a}^\mathrm{true}_\mathrm{LS} + \Gamma_\mathrm{SS} \mathbf{a}^\mathrm{true}_\mathrm{SS} - \Gamma_\mathrm{LS} \mathbf{a}_\mathrm{LS} \right) = 0,
\end{equation}
where~$\nabla_{\mathbf{a}_\mathrm{LS}}$ is an operator whose elements are derivatives with respect to each individual large-scale~$a_{\ell m}$.
The use of the Moore-Penrose inverse immediately returns equation~\eqref{eq:shift_least_square} since by construction~$\Gamma^+_\mathrm{LS} \Gamma_\mathrm{LS}=I_{N_\mathrm{LS}}$, with~$I_n$ being the identity matrix of dimension~$n$.

For each value of~$\ell$, the vector~$\mathbf{a}_\ell$ is constructed by introducing a projection matrix~$\mathbb{P}_\ell$ to isolate the~$(2\ell+1)$ components of~$\mathbf{a}_\mathrm{LS}$ corresponding to~$a_{\ell m}$.
In practice, $\mathbb{P}_\ell$ is a rectangular matrix with dimensions~$(2\ell+1)\times N_\mathrm{LS}$ defined as
\begin{equation}
    \mathbb{P}_\ell = 
    \left( \begin{matrix}
        0_{(2\ell+1) \times (\ell^2-4)} & I_{(2\ell+1)} & 0_{(2\ell+1 ) \times (N_\mathrm{LS}-\ell^2-2\ell+3)}
    \end{matrix} \right),
\end{equation}
where~$0_{i\times j}$ is a rectangular matrix where all elements are zeros.
For instance, we have that
\begin{equation}
    \mathbb{P}_2 = 
    \left( \begin{matrix}
    I_5 & 0_{5\times N_\mathrm{LS}-5}
    \end{matrix} \right), \quad
    \mathbb{P}_3 = 
    \left( \begin{matrix}
    0_{7\times 5} & I_7 & 0_{7\times N_\mathrm{LS}-12}
    \end{matrix} \right), \quad 
    \mathbb{P}_4 = 
    \left( \begin{matrix}
    0_{9\times 12} & I_9 & 0_{7\times N_\mathrm{LS}-21}
    \end{matrix} \right),\quad \mathrm{etc.},
\end{equation}
and, for the purpose of developing future intuition, we note that the square matrix~$\mathbb{P}^T_\ell \mathbb{P}_\ell$ is given by
\begin{equation}
    \mathbb{P}^T_\ell \mathbb{P}_\ell = 
    \left( \begin{matrix}
    0_{(\ell^2-4)\times (\ell^2-4)} & 0_{(\ell^2-4)\times (2\ell+1)} & 0_{(\ell^2-4)\times (N_\mathrm{LS}-\ell^2-2\ell+3)} \\
    0_{(2\ell+1)\times(\ell^2-4)} & I_{(2\ell+1)} & 0_{(2\ell+1)\times(N_\mathrm{LS}-\ell^2-2\ell+3)} \\ 
    0_{(N_\mathrm{LS}-\ell^2-2\ell+3) \times (\ell^2-4)} & 0_{(N_\mathrm{LS}-\ell^2-2\ell+3) \times (2\ell+1)} & 0_{(N_\mathrm{LS}-\ell^2-2\ell+3) \times (N_\mathrm{LS}-\ell^2-2\ell+3)} \\
    \end{matrix} \right).
    \label{eq:Pell_def}
\end{equation}

The convoluted matrix product of equation~\eqref{eq:cl_leakage_fitting} can be simplified by defining a ``leakage'' matrix
\begin{equation}
    \mathcal{M}_\ell = \Gamma^T_\mathrm{SS} \left( \Gamma^+_\mathrm{LS}\right)^T \mathbb{P}^T_\ell \mathbb{P}_\ell \Gamma^+_\mathrm{LS} \Gamma_\mathrm{SS},
\end{equation}
in such a way that
\begin{equation}
    \begin{aligned}
        C^\mathrm{leakage}_\ell &= \frac{1}{2\ell+1} \left\langle \left( \mathbf{a}^\mathrm{true}_\mathrm{SS} \right)^T \mathcal{M}_\ell \mathbf{a}^\mathrm{true}_\mathrm{SS} \right\rangle = \frac{1}{2\ell+1} \sum_{ij} \left\langle \mathbf{a}^\mathrm{true}_{\mathrm{SS},i} \mathcal{M}_{\ell,ij} \mathbf{a}^\mathrm{true}_{\mathrm{SS},j} \right\rangle \\
        &= \frac{1}{2\ell+1} \sum_{\ell' m'} \sum_{\ell'' m''} \left\langle a^\mathrm{true}_{\ell' m'} \mathcal{M}_{\ell,i_{\ell' m'} j_{\ell'' m''}} a^\mathrm{true}_{\ell'' m''} \right\rangle = \frac{1}{2\ell+1} \sum_{\ell' m'} \sum_{\ell'' m''} \mathcal{M}_{\ell, i_{\ell' m'} j_{\ell'' m''}} \delta^K_{\ell' \ell''} \delta^K_{m' m''} C^\mathrm{true}_{\ell'}  \\
        &= \sum_{\ell'} \left( \frac{1}{2\ell+1} \sum_{m'} \mathcal{M}_{\ell, i_{\ell' m'} i_{\ell' m'}} \right) C^\mathrm{true}_{\ell'} = \sum_{\ell'} M_{\ell\ell'} C^\mathrm{true}_{\ell'},
    \end{aligned}
\label{eq:clleakage_explicitcalculation}
\end{equation}
where we indicate with~$i_{\ell m}$ the matrix row/column index corresponding to the~$(\ell,m)$ pair of indices.

\begin{figure}[ht]
    \centerline{
    \includegraphics[width=\columnwidth]{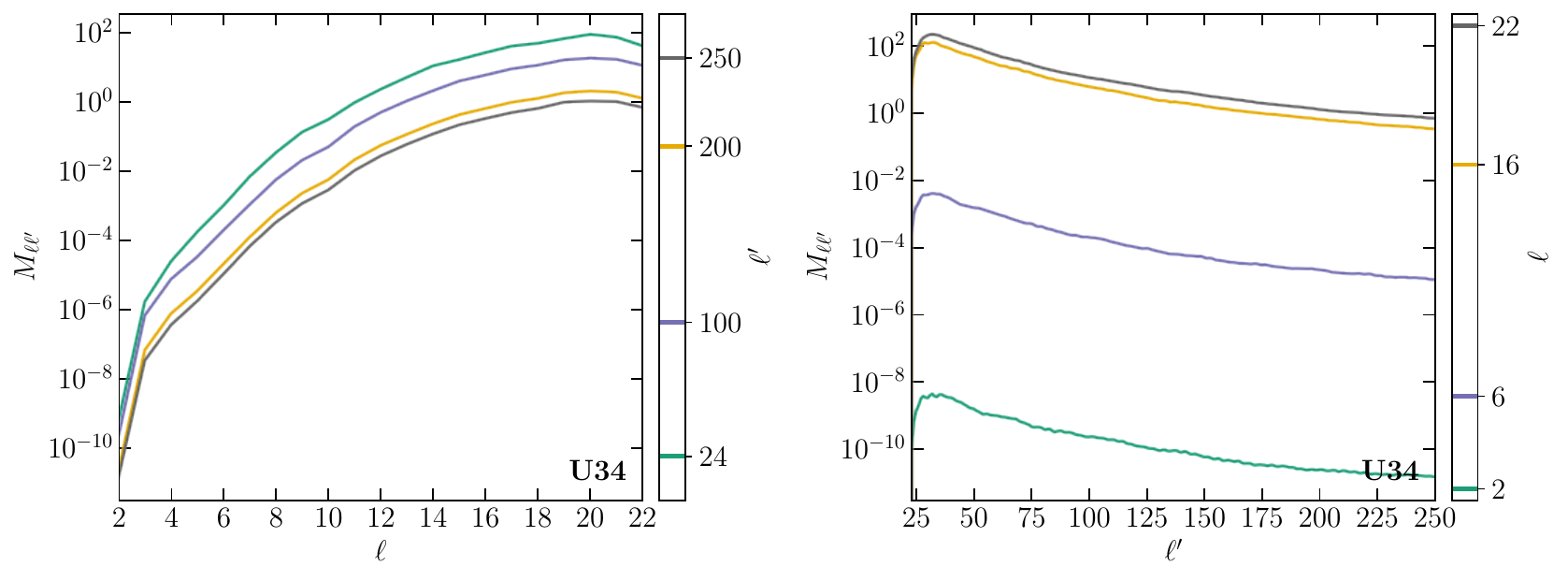}}
    \centerline{
    \includegraphics[width=\columnwidth]{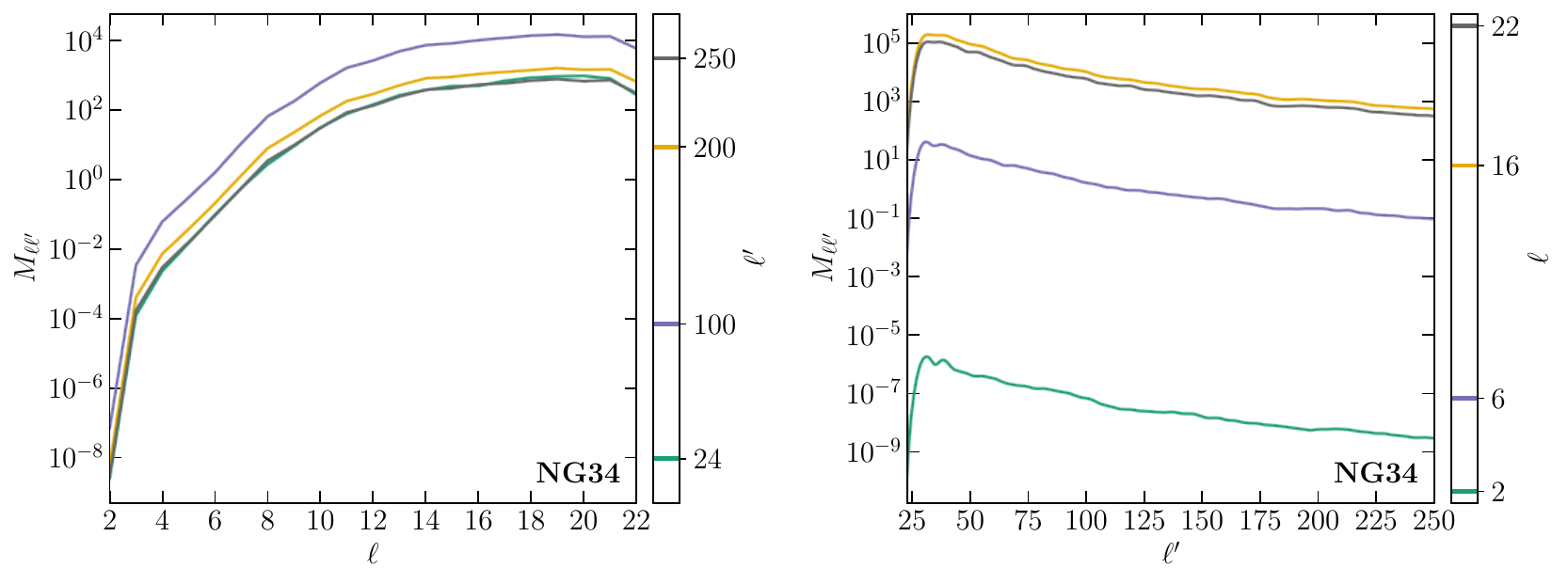}}
    \caption{Value of the mode-mixing function~$M_{\ell \ell'}$ in term of large-scale~$\ell$ and small-scale~$\ell'$ multipoles for the U34 (\textit{top panels}) and NG34 (\textit{bottom panels}) pulsar configurations.}
\label{fig:Mllprime_UNG34}
\end{figure}

We show in figure~\ref{fig:Mllprime_UNG34} the structure of the mode-mixing function for the pulsar configurations U34 and NG34 in the top and bottom panels, respectively.
By comparing this figure with the I34 pulsar case shown in figure~\ref{fig:Mllprime_pure_geometry}, the overall shape of this function does not vary significantly between different pulsar geometries.
However, there is a significant difference in terms of amplitude of~$M_{\ell \ell'}$, likely introduced by the presence of nearly singular pulsar configurations.
Therefore, we expect that the small-scale leakage in these cases is considerably larger than that obtained for the I34 configuration.

%%%%%%%%%%%%%%%%%%%%%%%%%%%%%%%%%%%%%%%%%%%%%%%%%%%%%%%%%%%%%%%%%%%%%%%%%%%%%%%%%%%%%%%%%%%%%%%%%%%%%%%%%%%%%%%%%%%%%%%%%%%%%%%%%%%%%%%%%%

\subsection{Statistical intuition}
Similarly to the purely geometric case, the position of the maximum of the likelihood is given by
\begin{equation}
    \nabla_{\mathbf{a}_\mathrm{LS}} \log \mathcal{L} \simeq \Gamma^T_\mathrm{LS} \mathrm{Cov}^{-1} \left( \boldsymbol{\RR} - \Gamma_\mathrm{LS} \mathbf{a}_\mathrm{LS} \right) = 0.
\end{equation}
In this case, the normal equations is commonly recast as
\begin{equation}
    \mathcal{H} \mathbf{a}_\mathrm{LS} = \mathbf{v},
    \label{eq:normal equations}
\end{equation}
where the~$\mathcal{H} = \Gamma^T_\mathrm{LS} \mathrm{Cov}^{-1} \Gamma_\mathrm{LS} \simeq \nabla^2_{\mathbf{a}_\mathrm{LS}} \log \mathcal{L}$ is proportional to the likelihood Hessian matrix, and~$\mathbf{v}~= \Gamma^T_\mathrm{LS} \mathrm{Cov}^{-1} \boldsymbol{\RR}$ is the ``likelihood score''.
As before, it is straightforward to recover equation~\eqref{eq:shift_maximum_likelihood} from the set of normal equations.
In this case, the bias introduced depends not only on the pulsar geometry but also on how the covariance matrix is modeled.
This result is totally analogous of what in Cosmology is typically referred to as ``bestfit shift'' due to an incorrect modeling of the physical observable, see, e.g., ref.~\cite{Bernal_2020} and refs. therein.

Finally, we observe that the algebra underlying equation~\eqref{eq:cl_leakage_likelihood} is completely equivalent to that leading to equation~\eqref{eq:clleakage_explicitcalculation}, except for using as leakage matrix
\begin{equation}
    \mathcal{M}'_\ell = \left[ \Gamma^{-1}_\mathrm{LS} \mathrm{Cov}^{-1} \Gamma_\mathrm{SS} \right]^T \left[ \left(\Gamma^T_\mathrm{LS} \mathrm{Cov}^{-1} \Gamma_\mathrm{LS} \right)^{-1} \right]^T \mathbb{P}^T_\ell \mathbb{P}_\ell \left(\Gamma^T_\mathrm{LS} \mathrm{Cov}^{-1} \Gamma_\mathrm{LS} \right)^{-1} \Gamma^{-1}_\mathrm{LS} \mathrm{Cov}^{-1} \Gamma_\mathrm{SS},
\end{equation}
leading to an analogous definition of a matrix~$M'_{\ell \ell'}$.

%%%%%%%%%%%%%%%%%%%%%%%%%%%%%%%%%%%%%%%%%%%%%%%%%%%%%%%%%%%%%%%%%%%%%%%%%%%%%%%%%%%%%%%%%%%%%%%%%%%%%%%%%%%%%%%%%%%%%%%%%%%%%%%%%%%%%%%%%%

\subsection{Covariance matrix}

In this section we compute for the first time the form of the covariance matrix for anisotropies searches without discarding the intrinsic cosmological signal, along the lines of what described in ref.~\cite{alihaimoud2020}.
For the sake of simplicity we consider a single, narrow frequency bin; however, the derivation presented here can easily being generalized to the case of multiple, wide frequency bins.
Suppose to define for each pulsar an estimator of the measured timing residual in equation~\eqref{eq:measured_timing_residual} as
\begin{equation}
    \hat{R}^\mathrm{meas}_p = \hat{R}_p + \hat{n}_p,
\end{equation}
so that the correlated residual estimator simply reads as
\begin{equation}
    \widehat{\RR}^\mathrm{meas}_{pq} = \hat{R}^\mathrm{meas}_p \hat{R}^\mathrm{meas}_q = \left( \hat{R}_p + \hat{n}_p \right) \left( \hat{R}_q + \hat{n}_q \right). 
\end{equation}
This estimator is built is such a way that its expectation value over the source and noise properties is given by
\begin{equation}
    \left\langle \widehat{\RR}^\mathrm{meas}_{pq} \right\rangle_{s,n} = \RR^\mathrm{iso}_{pq} + \RR^\mathrm{anis}_{pq} + \delta^K_{pq} \sigma^2_p,
\end{equation}
where, as already commented, the anisotropic component still has a residual degree of stochasticity such that
\begin{equation}
    \left\langle \RR^\mathrm{anis}_{pq} \right\rangle_p = \sum_{\ell m} \Gamma^{pq}_{\ell m} \left\langle a_{\ell m} \right\rangle_p = 0, 
\end{equation}
and
\begin{equation}
    \left\langle \RR^\mathrm{anis}_{pq} \RR^\mathrm{anis}_{rs} \right\rangle_p = \sum_{\ell m} \sum_{\ell' m'} \Gamma^{pq}_{\ell m} \Gamma^{rs}_{\ell' m'} \left\langle a_{\ell m} a_{\ell' m'} \right\rangle_p = \sum_{\ell m} \sum_{\ell' m'} \Gamma^{pq}_{\ell m} \Gamma^{rs}_{\ell m} C_\ell. 
\end{equation}
In this section we use the angle bracket subscripts~$(s,n,p)$ to indicate ensemble averages over source, noise and position properties, respectively, and should not be confuse with the pulsar indices.
In particular, the angle bracket subscript~$p$ has been omitted in the main text for simplicity; however, in this section it is useful to keep track about what sort of averages we are implementing in different stages of the derivation.

Therefore, it is immediate to construct an estimator of the measured anisotropic correlated residual as
\begin{equation}
\label{eq:Rpq_anis_meas}
    \widehat{\RR}^\mathrm{anis,meas}_{pq} = \widehat{\RR}^\mathrm{meas}_{pq} - \RR^\mathrm{iso}_{pq} - \delta^K_{pq} \sigma^2_p = \hat{R}^\mathrm{meas}_p \hat{R}^\mathrm{meas}_q - \RR^\mathrm{iso}_{pq} - \delta^K_{pq} \sigma^2_p,
\end{equation}
where, as expected, 
\begin{equation}
\label{eq:Rpq_anis_meas_props}
    \left\langle \widehat{\RR}^\mathrm{anis,meas}_{pq} \right\rangle_{s,n} = \RR^\mathrm{anis,th}_{pq}, \qquad \left\langle \widehat{\RR}^\mathrm{anis,meas}_{pq} \right\rangle_{s,n,p} = 0,
\end{equation}
and in this case the ``theoretical'' correlated residual~$\RR^\mathrm{anis,th}_{pq}$ has the meaning of the correlated residual of the realization of the sky we observe.
At this point, the~$(i_{pq},j_{rs})$ element of the covariance matrix read as
\begin{equation}
    \begin{aligned}
        \mathrm{Cov}_{i_{pq}j_{rs}} &= \left\langle \left[ \widehat{\RR}^\mathrm{anis,meas}_{pq} - \RR^\mathrm{anis,th}_{pq} \right] \left[ \widehat{\RR}^\mathrm{anis,meas}_{rs} - \RR^\mathrm{anis,th}_{rs} \right] \right\rangle_{s,n,p} \\
        &= \left\langle \left[ \hat{R}^\mathrm{meas}_p \hat{R}^\mathrm{meas}_q - \RR^\mathrm{iso}_{pq} - \delta^K_{pq} \sigma^2_p - \RR^\mathrm{anis,th}_{pq} \right] \left[ \hat{R}^\mathrm{meas}_r \hat{R}^\mathrm{meas}_s - \RR^\mathrm{iso}_{rs} - \delta^K_{rs} \sigma^2_r - \RR^\mathrm{anis,th}_{rs} \right] \right\rangle_{s,n,p}.
    \end{aligned}
\end{equation}
Given the statistical properties described before, it is straightforward to evaluate all the individual terms appearing in this equation.
In particular, we have
\begin{equation}
    \begin{aligned}
        \left\langle \hat{R}^\mathrm{meas}_p \hat{R}^\mathrm{meas}_q \hat{R}^\mathrm{meas}_r \hat{R}^\mathrm{meas}_s \right\rangle_{s,n,p} &= \left\langle \left( \hat{R}_p + \hat{n}_p \right) \left( \hat{R}_q + \hat{n}_q \right) \left( \hat{R}_r + \hat{n}_r \right) \left( \hat{R}_s + \hat{n}_s \right) \right\rangle_{s,n,p} \\
        &= \Big\langle \hat{R}_p \hat{R}_q \hat{R}_r \hat{R}_s  + \left( \hat{R}_p \hat{R}_q \hat{R}_r \hat{n}_s + \mathrm{3\ perm.} \right) \\
        &\qquad\quad + \left( \hat{R}_p \hat{R}_q \hat{n}_r \hat{n}_s + \mathrm{5\ perm.} \right) + \left( \hat{R}_p \hat{n}_q \hat{n}_r \hat{n}_s + \mathrm{3\ perm.} \right) \\
        &\qquad\qquad\qquad + \hat{n}_p \hat{n}_q \hat{n}_r \hat{n}_s \Big\rangle_{s,n,p} \\
        &= \left(\RR^\mathrm{iso}_{pq} + \delta^K_{pq} \sigma^2_p \right) \left( \RR^\mathrm{iso}_{rs} + \delta^K_{rs} \sigma^2_r \right) \\
        &\qquad + \left( \RR^\mathrm{iso}_{pr} + \delta^K_{pr} \sigma^2_p \right) \left( \RR^\mathrm{iso}_{qs} + \delta^K_{qs} \sigma^2_q \right) \\
        &\qquad\qquad + \left( \RR^\mathrm{iso}_{ps} + \delta^K_{ps} \sigma^2_p \right) \left( \RR^\mathrm{iso}_{qr} + \delta^K_{qr} \sigma^2_q \right) \\
        &\qquad\qquad\qquad + \sum_{\ell m} \left[ \Gamma^{pq}_{\ell m} \Gamma^{rs}_{\ell m} + \Gamma^{pr}_{\ell m} \Gamma^{qs}_{\ell m} + \Gamma^{ps}_{\ell m} \Gamma^{qr}_{\ell m} \right] C_\ell,
    \end{aligned}
\end{equation}
where we used the fact that the only non-vanishing terms are
\begin{equation}
    \begin{aligned}
        \left\langle \hat{R}_p \hat{R}_q \hat{R}_r \hat{R}_s \right\rangle_{s,n,p} &= \left\langle \left\langle \hat{R}_p \hat{R}_q \right\rangle_s \left\langle \hat{R}_r \hat{R}_s \right\rangle_s + \left\langle \hat{R}_p \hat{R}_r \right\rangle_s \left\langle \hat{R}_q \hat{R}_s \right\rangle_s + \left\langle \hat{R}_p \hat{R}_s \right\rangle_s \left\langle \hat{R}_q \hat{R}_r \right\rangle_s\right\rangle_{p} \\
        &= \left\langle \left( \RR^\mathrm{iso}_{pq} + \widehat{\RR}^\mathrm{anis}_{pq} \right) \left( \RR^\mathrm{iso}_{rs} + \widehat{\RR}^\mathrm{anis}_{rs} \right) + \left( \RR^\mathrm{iso}_{pr} + \widehat{\RR}^\mathrm{anis}_{pr} \right) \left( \RR^\mathrm{iso}_{qs} + \widehat{\RR}^\mathrm{anis}_{qs} \right) \right. \\
        &\qquad\qquad\qquad \left. + \left( \RR^\mathrm{iso}_{ps} + \widehat{\RR}^\mathrm{anis}_{ps} \right) \left( \RR^\mathrm{iso}_{qr} + \widehat{\RR}^\mathrm{anis}_{qr} \right) \right\rangle_{p} \\
        &= \RR^\mathrm{iso}_{pq} \RR^\mathrm{iso}_{rs} + \RR^\mathrm{iso}_{pr} \RR^\mathrm{iso}_{qs} + \RR^\mathrm{iso}_{ps} \RR^\mathrm{iso}_{qr} \\
        &\qquad\qquad + \sum_{\ell m} \left[ \Gamma^{pq}_{\ell m} \Gamma^{rs}_{\ell m} + \Gamma^{pr}_{\ell m} \Gamma^{qs}_{\ell m} + \Gamma^{ps}_{\ell m} \Gamma^{qr}_{\ell m} \right] C_\ell,\\
        \left\langle \hat{n}_p \hat{n}_q \hat{n}_r \hat{n}_s \right\rangle_{s,n,p} &= \left\langle \left\langle \hat{n}_p \hat{n}_q \right\rangle_n \left\langle \hat{n}_r \hat{n}_s \right\rangle_n + \left\langle \hat{n}_p \hat{n}_r \right\rangle_n \left\langle \hat{n}_q \hat{n}_s \right\rangle_n + \left\langle \hat{n}_p \hat{n}_s \right\rangle_n \left\langle \hat{n}_q \hat{n}_r \right\rangle_n \right\rangle_{s,p} \\
        &= \left\langle \delta^K_{pq} \delta^K_\mathrm{rs} \sigma^2_p \sigma^2_r + \delta^K_{pr} \delta^K_\mathrm{qs} \sigma^2_p \sigma^2_q + \delta^K_{ps} \delta^K_\mathrm{qr} \sigma^2_p \sigma^2_q \right\rangle_{s,p} \\
        &= \delta^K_{pq} \delta^K_\mathrm{rs} \sigma^2_p \sigma^2_r + \delta^K_{pr} \delta^K_\mathrm{qs} \sigma^2_p \sigma^2_q + \delta^K_{ps} \delta^K_\mathrm{qr} \sigma^2_p \sigma^2_q, \\
        \left\langle \hat{R}_p \hat{R}_q \hat{n}_r \hat{n}_s \right\rangle_{s,n,p} &= \left\langle \left\langle \hat{R}_p \hat{R}_q \right\rangle_s \left\langle \hat{n}_r \hat{n}_s \right\rangle_n \right\rangle_{p} = \RR^\mathrm{iso}_{pq} \delta^K_{rs} \sigma^2_r,
    \end{aligned}
\end{equation}
and permutations of the latter.
Similarly, we have that
\begin{equation}
    \begin{aligned}
        \left\langle \hat{R}^\mathrm{meas}_p \hat{R}^\mathrm{meas}_q \right. & \left. \left( \RR^\mathrm{iso}_{rs} + \delta^K_{rs} \sigma^2_r + \RR^\mathrm{anis,th}_{rs} \right) \right\rangle_{s,n,p} = \left\langle \left( \RR^\mathrm{iso}_{pq} + \delta^K_{pq} \sigma^2_p + \RR^\mathrm{anis,th}_{pq} \right) \hat{R}^\mathrm{meas}_r \hat{R}^\mathrm{meas}_s \right\rangle_{s,n,p} \\
        &= \left\langle \left( \RR^\mathrm{iso}_{pq} + \delta^K_{pq} \sigma^2_p + \RR^\mathrm{anis,th}_{pq} \right) \left( \RR^\mathrm{iso}_{rs} + \delta^K_{rs} \sigma^2_r + \RR^\mathrm{anis,th}_{rs} \right) \right\rangle_{s,n,p} \\
        &= \left( \RR^\mathrm{iso}_{pq} + \delta^K_{pq} \sigma^2_p \right) \left( \RR^\mathrm{iso}_{rs} + \delta^K_{rs} \sigma^2_r \right) + \sum_{\ell m} \Gamma^{pq}_{\ell m} \Gamma^{rs}_{\ell m} C_\ell.
    \end{aligned}
\end{equation}
Therefore, it is immediate to recover that the elements of the covariance matrix read as in equation~\eqref{eq:shift_maximum_likelihood}, when the noise is absent, or as in equation~\eqref{eq:full_cov}, when the noise is present.

%%%%%%%%%%%%%%%%%%%%%%%%%%%%%%%%%%%%%%%%%%%%%%%%%%%%%%%%%%%%%%%%%%%%%%%%%%%%%%%%%%%%%%%%%%%%%%%%%%%%%%%%%%%%%%%%%%%%%%%%%%%%%%%%%%%%%%%%%%

%\input{000_app_IFFF}

%%%%%%%%%%%%%%%%%%%%%%%%%%%%%%%%%%%%%%%%%%%%%%%%%%%%%%%%%%%%%%%%%%%%%%%%%%%%%%%%%%%%%%%%%%%%%%%%%%%%%%%%%%%%%%%%%%%%%%%%%%%%%%%%%%%%%%%%%%

%%%%%%%%%%%%%%%%%%%%%%%%%%%%%%%%%%%%%%%%%%%%%%%%%%%%%%%%%%%%%%%%%%%%%%%%%%%%%%%%%%%%%%%%%%%%%%%%%%%%%%%%%%%%%%%%%%%%%%%%%%%%%%%%%%%%%%%%%%

\section{Regularization schemes}
\label{app:regularization}

%%%%%%%%%%%%%%%%%%%%%%%%%%%%%%%%%%%%%%%%%%%%%%%%%%%%%%%%%%%%%%%%%%%%%%%%%%%%%%%%%%%%%%%%%%%%%%%%%%%%%%%%%%%%%%%%%%%%%%%%%%%%%%%%%%%%%%%%%%

\subsection{Likelihood regularization}

After the introduction of the ridge regularization factor, the normal equations read as
\begin{equation}
    \left[\mathcal{H} + \lambda \mathcal{P} \right] \mathbf{a}_\mathrm{LS} = \mathbf{v} = \mathcal{H} \mathbf{a}^\mathrm{true}_\mathrm{LS} + \Gamma^T_\mathrm{LS} \mathrm{Cov}^{-1} \Gamma_\mathrm{SS} \mathbf{a}^\mathrm{true}_\mathrm{SS},
\end{equation}
therefore, the explicit form of equation~\eqref{eq:maximumlikelihoodshift_regularization} is immediately recovered by substituting~$\mathcal{H} = (\mathcal{H}+\lambda\mathcal{P}) - \lambda\mathcal{P}$ in the RHS of this equation.
For the purpose of this work, we choose as penalty matrix~$\mathcal{P} = \mathrm{med}_H \times I_{N_\mathrm{LS}}$, where~$\mathrm{med}_H$ is median value of the Hessian matrix diagonal.
For~$\lambda=1$, this coefficient introduces a regularization with an order of magnitude comparable to that of the ill-conditioned Hessian matrix.

As noted previously, it is possible to infer the phenomenology due to the introduction of a regularization scheme on the~$\chi^2_\mathrm{LMS}$ statistics, i.e., in the map-making scenario, by substituting~$\mathrm{Cov} = I_{N_\mathrm{LS}}$ in equation~\eqref{eq:maximumlikelihoodshift_regularization}.
In that case, we have that
\begin{equation}
    \mathbf{a}_\mathrm{LS} = \left[ I - \left(\Gamma^T_\mathrm{LS} \Gamma_\mathrm{LS} + \lambda\mathcal{P} \right)^{-1} \lambda\mathcal{P} \right] \mathbf{a}^\mathrm{true}_\mathrm{LS} + \left(\Gamma^T_\mathrm{LS} \Gamma_\mathrm{LS} + \lambda\mathcal{P} \right)^{-1} \Gamma^T_\mathrm{LS} \Gamma_\mathrm{SS} \mathbf{a}^\mathrm{true}_\mathrm{SS}.
\end{equation}
Therefore, also in this case, the biased angular power spectra have an identical structure as those obtained via the likelihood approach discussed in section~\ref{subsec:regularization_scheme}.
Finally, even if we present only the case of ridge regularization, other schemes as, for instance, SVD regularization, can be recast as a special ridge regularization case, thus the generality of this formulation and of our conclusions.

%%%%%%%%%%%%%%%%%%%%%%%%%%%%%%%%%%%%%%%%%%%%%%%%%%%%%%%%%%%%%%%%%%%%%%%%%%%%%%%%%%%%%%%%%%%%%%%%%%%%%%%%%%%%%%%%%%%%%%%%%%%%%%%%%%%%%%%%%%

\subsection{Covariance matrix regularization}

In the case of regularizing the covariance matrix, we implement a SVD (Single Value Decomposition) regularization scheme.
First, we perform an eigendecomposition of the covariance matrix as~$\mathrm{Cov} = U \Sigma U^{-1}$, where~$U$ is the orthogonal matrix of eigenvectors, and~$\Sigma=\mathrm{diag}(e_j)$ is a diagonal matrix containing the eigenvalues~$\lbrace e_j \rbrace$ of the covariance.
The regularized version of the inverse of the covariance is then explicitly computed as
\begin{equation}
    \mathrm{Cov}^{-1}_\mathrm{reg} = U \Sigma^{-1}_\mathrm{reg} U^{-1},
\end{equation}
where the elements of the diagonal matrix~$\Sigma^{-1}_\mathrm{reg}$ read as 
\begin{equation}
    \left( \Sigma^{-1}_\mathrm{reg} \right)_{jj} = 
    \left\lbrace \begin{matrix}
        e^{-1}_j, & \qquad e_j \geq e_\mathrm{thres}, \\
        0, & \qquad e_j < e_\mathrm{thres}, \\
    \end{matrix} \right.
    \label{eq:SVD_def}
\end{equation}
and~$e_\mathrm{thres}$ is an arbitrary threshold value.
When only this regularization scheme is implemented, the normal equations read as
\begin{equation}
    \mathcal{H}_\mathrm{reg} \mathbf{a}_\mathrm{LS} = \mathbf{v}_\mathrm{reg},
\end{equation}
where the regularized Hessian and likelihood score are given by~$\mathcal{H}_\mathrm{reg} = \Gamma^T_\mathrm{LS} \mathrm{Cov}^{-1}_\mathrm{reg} \Gamma_\mathrm{LS}$ and~$\mathbf{v}_\mathrm{reg} = \Gamma^T_\mathrm{LS} \mathrm{Cov}^{-1}_\mathrm{reg} \boldsymbol{\RR}$, respectively.

%%%%%%%%%%%%%%%%%%%%%%%%%%%%%%%%%%%%%%%%%%%%%%%%%%%%%%%%%%%%%%%%%%%%%%%%%%%%%%%%%%%%%%%%%%%%%%%%%%%%%%%%%%%%%%%%%%%%%%%%%%%%%%%%%%%%%%%%%%

\subsection{Combining regularization schemes}

\begin{figure}[ht]
    \centerline{
    \includegraphics[width=\columnwidth]{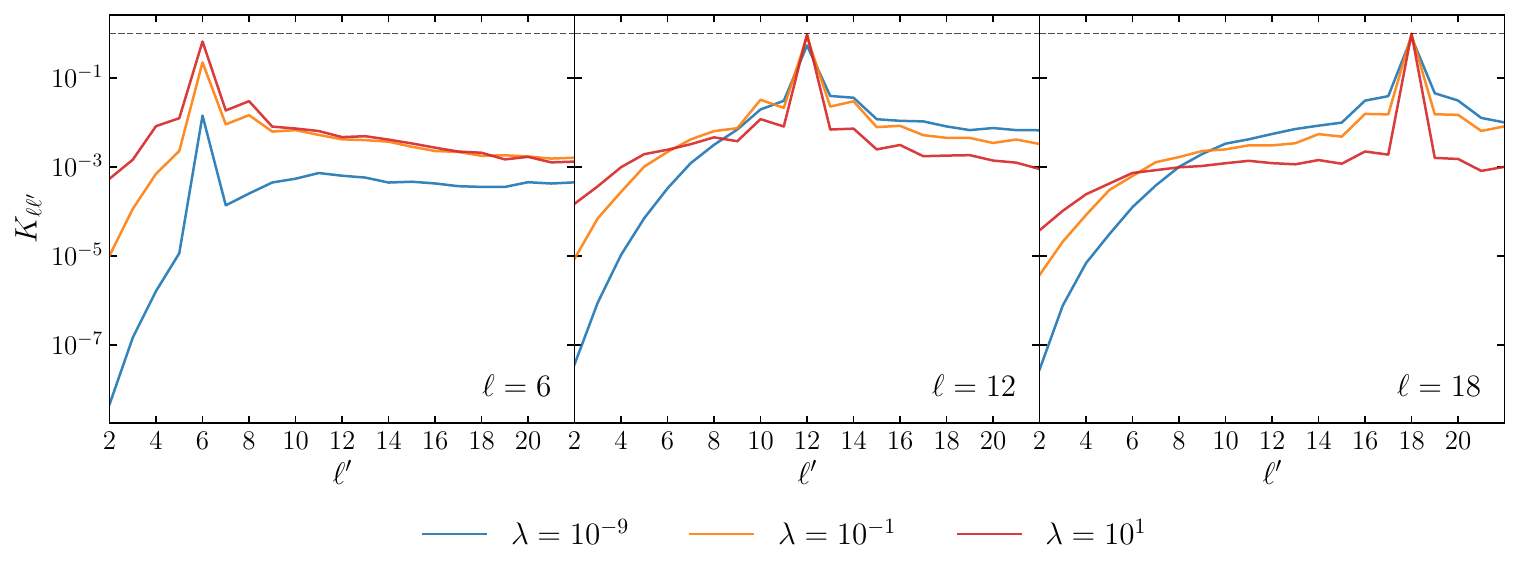}}
    \caption{Values of the regularization bias matrix for different $\ell=6,12,18$ large-scale multipoles (\textit{left, center,} and \textit{right panel}, respectively) for growing values of the regularization parameter~$\lambda$ in the NG34 pulsar configuration.
    }
\label{fig:Kllprime}
\end{figure}

When these two regulation schemes are simultaneously implemented, the estimate of the large-scale coefficients that maximize the likelihood is given by
\begin{equation}
    \mathbf{a}_\mathrm{LS} = \left[ I - \left(\mathcal{H}_\mathrm{reg} + \lambda\mathcal{P} \right)^{-1} \lambda\mathcal{P} \right] \mathbf{a}^\mathrm{true}_\mathrm{LS} + \left(\mathcal{H}_\mathrm{reg} + \lambda\mathcal{P} \right)^{-1} \Gamma^T_\mathrm{LS} \mathrm{Cov}^{-1}_\mathrm{reg} \Gamma_\mathrm{SS} \mathbf{a}^\mathrm{true}_\mathrm{SS},
\end{equation}
which now explicitly depends on the arbitrary choice of both~$\lambda$ and~$e_\mathrm{thres}$. 
In this case, we have to introduce a regularization bias matrix defined as
\begin{equation}
    \begin{aligned}
        \mathcal{K}_\ell &= \lambda \left[ \left[ \mathbb{P}^T_\ell \mathbb{P}_\ell \left(\mathcal{H}_\mathrm{reg} + \lambda\mathcal{P} \right)^{-1} \mathcal{P} \right]^T + \mathbb{P}^T_\ell \mathbb{P}_\ell \left(\mathcal{H}_\mathrm{reg} + \lambda\mathcal{P} \right)^{-1} \mathcal{P} \right] \\
        &\qquad\qquad\qquad - \lambda^2 \left[ \left(\mathcal{H}_\mathrm{reg} + \lambda\mathcal{P} \right)^{-1} \mathcal{P} \right]^T \mathbb{P}^T_\ell \mathbb{P}_\ell \left(\mathcal{H}_\mathrm{reg} + \lambda\mathcal{P} \right)^{-1} \mathcal{P}, \\
    \end{aligned}
\end{equation}
therefore, the bias introduced by ridge regularization is given by
\begin{equation}
    C^\mathrm{reg}_\ell = \frac{1}{2\ell+1} \left\langle \left( \mathbf{a}^\mathrm{true}_\mathrm{LS} \right)^T \mathcal{K}_\ell \mathbf{a}^\mathrm{true}_\mathrm{LS} \right\rangle = \sum_{\ell'} \left( \frac{1}{2\ell+1} \sum_{m'} \mathcal{K}_{\ell, i_{\ell' m'} i_{\ell' m'}} \right) C^\mathrm{true}_{\ell'} = \sum_{\ell'=2}^{\lmaxrec} K_{\ell\ell'} C^\mathrm{true}_{\ell'},
\end{equation}
where in this instance the sum runs over all the large-scale modes.

We show in figure~\ref{fig:Kllprime} the value of the regularization bias matrix for three different multipoles and ridge regularization parameters for the NG34 pulsar configuration.
These functions sharply peak around~$\ell'=\ell$; therefore in this scenario there is no strong large-scale mode coupling.
However, different geometries can potentially exhibit a stronger coupling, especially between nearby multipoles.
The same considerations apply also to the NG68 configuration, which is not shown for brevity.

\begin{figure}[ht]
    \centerline{
    \includegraphics[width=\columnwidth]{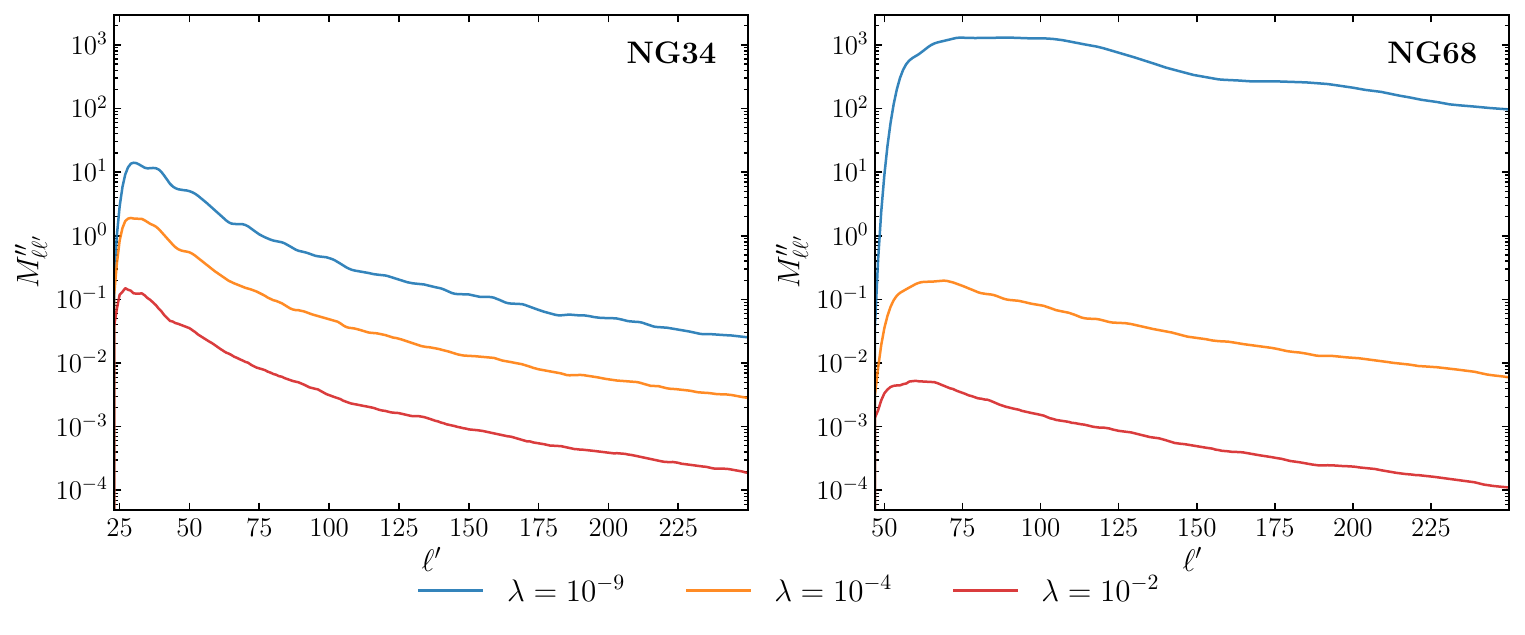}}
    \caption{Values of the mode-mixing function~$M''_{\ell\ell'}$ for the large-scale multipole~$\ell=18$ and growing values of the regularization parameter~$\lambda$, both for the NG34 (\textit{left panel}) and NG68 (\textit{right panel}) pulsar configurations.
    The magnitude of the function is strongly dependent on the presence of near perfect degeneracies which maximizes the leakage, especially in the NG68 case.
    }
\label{fig:Mllsecond}
\end{figure}

On the other hand, in this scenario, the leakage matrix reads as 
\begin{equation}
    \mathcal{M}''_\ell = \left[ \left(\mathcal{H}_\mathrm{reg} + \lambda\mathcal{P} \right)^{-1} \Gamma^T_\mathrm{LS} \mathrm{Cov}^{-1}_\mathrm{reg} \Gamma_\mathrm{SS} \right]^T \left(\mathcal{H}_\mathrm{reg} + \lambda\mathcal{P} \right)^{-1} \Gamma^T_\mathrm{LS} \mathrm{Cov}^{-1}_\mathrm{reg} \Gamma_\mathrm{SS},
\end{equation}
and it is responsible of an equivalent~$M''_{\ell \ell'}$ as described before.
We show in figure~\ref{fig:Mllsecond} how the mode-mixing matrix varies for growing values of the ridge regularization parameter, both for the NG34 and NG68 pulsar configurations.
In particular, we observe how the coupling between large- and small-scale modes is effectively diminished when the strength of the regularization increases. 
We also observe that in the presence of nearly-degenerate configurations, as for instance in NG68, coupling between different modes is significantly boosted. 
The trend of these functions is consistent across the range of large-scale multipoles and thus is not shown here for conciseness.

\begin{figure}[ht]
    \centerline{
    \includegraphics[width=\linewidth]{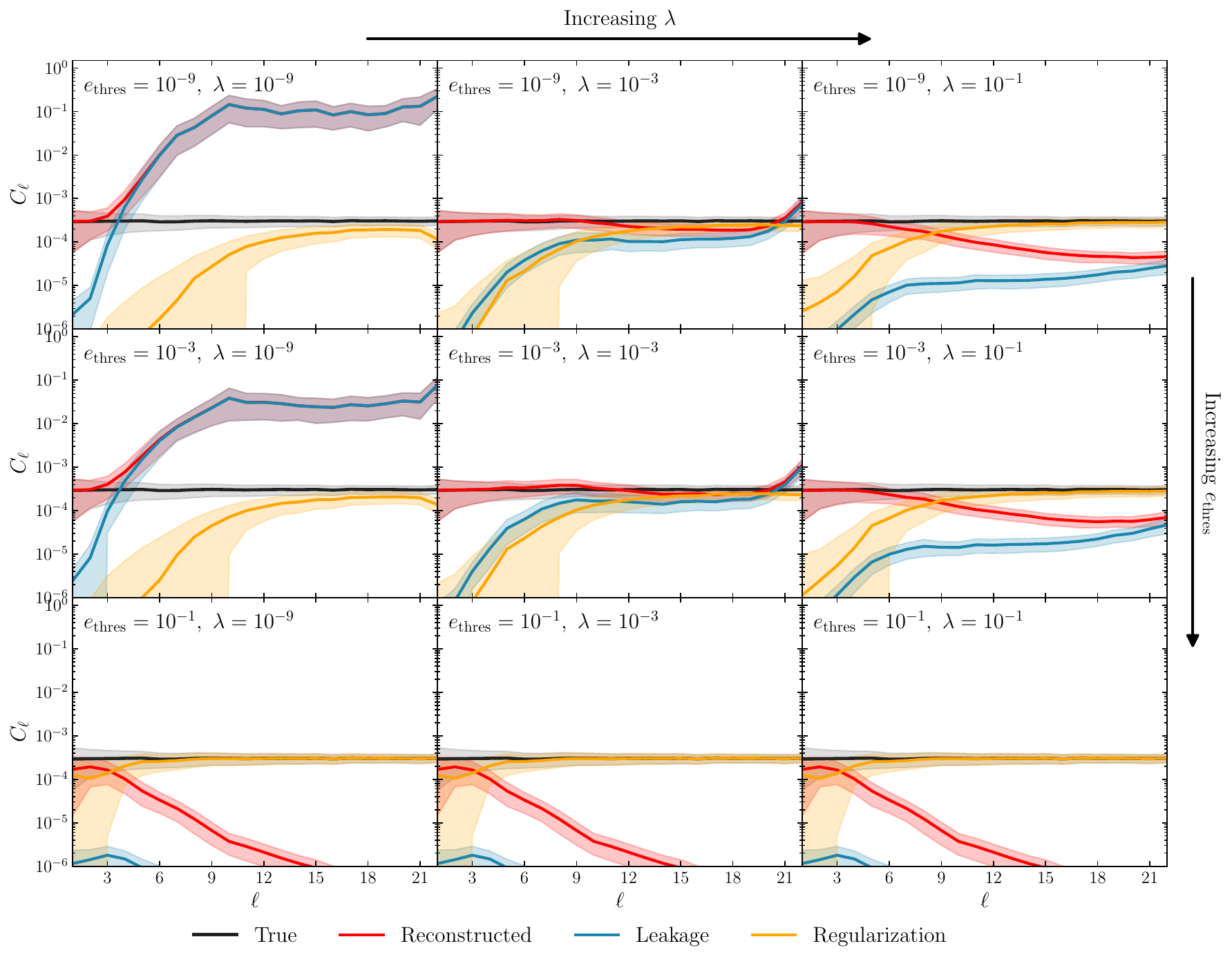}}
    \caption{Reconstructed angular power spectrum for the NG34 pulsar configuration for different values of the two regularization parameters, $\lambda$ (increasing from left to right) and~$e_\mathrm{thres}$ (increasing from top to bottom).}
\label{fig:regularization_parameters}
\end{figure}

Finally, we show in figure~\ref{fig:regularization_parameters} how the inferred angular power spectra change when both regularization parameters~$\lambda$ and~$e_\mathrm{thres}$ are varied independently of each other.
In all these cases, we note that there exists a subtle interplay between different choices for the regularization scheme that should be carefully evaluated in creating any anisotropy reconstruction algorithm.
Otherwise, the risk is to effectively remove power at small scales.
This behavior is replicated identically in the NG68 configuration.

%%%%%%%%%%%%%%%%%%%%%%%%%%%%%%%%%%%%%%%%%%%%%%%%%%%%%%%%%%%%%%%%%%%%%%%%%%%%%%%%%%%%%%%%%%%%%%%%%%%%%%%%%%%%%%%%%%%%%%%%%%%%%%%%%%%%%%%%%%

\bibliography{bibliography}
\bibliographystyle{utcaps}

\end{document}